\documentclass[12pt]{article}
\pdfoutput=1
\usepackage[latin1]{inputenc}

\usepackage{amsmath}
\usepackage{color}
\usepackage{amsfonts}
\usepackage{amssymb}
\usepackage{graphicx}
\usepackage{geometry}
\usepackage{amssymb,epsfig,subfigure}
\usepackage{hyperref}
\usepackage{comment}
\usepackage[font=footnotesize]{caption}

\makeatletter
\renewcommand\section{\@startsection {section}{1}{\z@}%
                                 {-3.5ex \@plus -1ex \@minus -.2ex}
                                   {2.3ex \@plus.2ex}%
                                   {\normalfont\large\bfseries}}
\renewcommand\subsection{\@startsection{subsection}{2}{\z@}%
                                   {-3.25ex\@plus -1ex \@minus -.2ex}%
                                     {1.5ex \@plus .2ex}%
                                     {\normalfont\bfseries}}
\renewcommand\subsubsection{\@startsection{subsubsection}{3}{\z@}%
                                   {-3.25ex\@plus -1ex \@minus -.2ex}%
                                     {1.5ex \@plus .2ex}%
                                     {\normalfont\itshape}}
\makeatother

\def\pplogo{\vbox{\kern-\headheight\kern -29pt
\halign{##&##\hfil\cr&{\ppnumber}\cr\rule{0pt}{2.5ex}&\ppdate\cr}}}
\makeatletter
\def\ps@firstpage{\ps@empty \def\@oddhead{\hss\pplogo}%
  \let\@evenhead\@oddhead 
}
\def\maketitle{\par
 \begingroup
 \def\thefootnote{\fnsymbol{footnote}}
 \def\@makefnmark{\hbox{$^{\@thefnmark}$\hss}}
 \if@twocolumn
 \twocolumn[\@maketitle]
 \else \newpage
 \global\@topnum\z@ \@maketitle \fi\thispagestyle{firstpage}\@thanks
 \endgroup
 \setcounter{footnote}{0}
 \let\maketitle\relax
 \let\@maketitle\relax
 \gdef\@thanks{}\gdef\@author{}\gdef\@title{}\let\thanks\relax}
\makeatother

\numberwithin{equation}{section}

\newcommand\eea{\end{eqnarray}}
\newcommand\bea{\begin{eqnarray}}

\def\beq{\begin{equation}}
\def\eeq{\end{equation}}

\newcommand{\be}{\begin{equation}}
\newcommand{\ee}{\end{equation}}
\newcommand{\ba}{\begin{align}}
\newcommand{\ea}{\end{align}}
\newcommand{\bg}{\begin{gather}}
\newcommand{\eg}{\end{gather}}
\newcommand{\bseq}{\begin{subequations}}
\newcommand{\eseq}{\end{subequations}}

\renewcommand{\tanh}{\mathop{\rm th}\nolimits}

\renewcommand{\t}{\tilde}

\newcommand{\tr}{{\rm tr}}
\newcommand{\mc}{\mathcal}

\begin{document}
\setcounter{page}0
\def\ppnumber{\vbox{\baselineskip14pt
}}
\def\ppdate{
} \date{}

\author{Horacio Casini, Ignacio Salazar Landea, Gonzalo Torroba\\
[7mm] \\
{\normalsize \it Centro At\'omico Bariloche and CONICET}\\
{\normalsize \it S.C. de Bariloche, R\'io Negro, R8402AGP, Argentina}
}

\bigskip
\title{\bf  The $g$-theorem and quantum \\ information theory
\vskip 0.5cm}
\maketitle

\begin{abstract}
We study boundary renormalization group flows between boundary conformal field theories in $1+1$ dimensions using methods of quantum information theory.
We define an entropic $g$-function for theories with impurities in terms of the relative entanglement entropy, and we prove that this $g$-function decreases along boundary renormalization group flows. This entropic $g$-theorem is valid at zero temperature, and is independent from the $g$-theorem based on the thermal partition function. We also discuss the mutual information in boundary RG flows, and how it encodes the correlations between the impurity and bulk degrees of freedom. Our results provide a quantum-information understanding of (boundary) RG flow as increase of distinguishability between the UV fixed point and the theory along the RG flow.
\end{abstract}
\bigskip

\newpage

\tableofcontents

\vskip 1cm

\section{Introduction}\label{sec:intro}

Quantum impurities and defects play an important role in different areas of theoretical physics, including condensed matter physics, gauge theories, and string theory. In order to understand the possible quantum field theories with defects and their dynamics, a key step is to classify boundary conditions that preserve some conformal invariance in bulk conformal field theories (CFTs), together with the renormalization group flows between different boundary conditions.

The best understood situation arises in two-dimensional CFTs with conformal boundaries, which led to the development of boundary CFT (BCFT). This case is especially interesting, because it arises from spherically symmetric magnetic impurities in metals (as in the famous Kondo problem  \cite{Kondo}) and also describes D-branes in string theory  \cite{Polchinski}. In a 2d CFT, Cardy found that a conformal boundary corresponds to a boundary state \cite{Cardy:1989ir}. Affleck and Ludwig defined a
$g$-function in terms of the difference between the thermal entropy with and without impurity, and used the formalism of boundary states to compute it \cite{Affleck:1991tk}. The boundary entropy plays the role of a ground-state degeneracy associated to the impurity, and these authors conjectured that $g$ decreases under renormalization. A key result in this direction is the proof of Friedan and Konechny  that establishes that $g$ indeed decreases monotonically along boundary RG flows \cite{Friedan:2003yc}.

A crucial property of the boundary entropy is that its value at a fixed point is in fact (part of) an entanglement entropy, as shown in \cite{Calabrese:2009qy}. However, this equivalence is not valid away from fixed points, as it uses the conformal map between the plane and the cylinder in two dimensions. This raises the important question of whether there exists an ``entropic $g$-function'' that decreases monotonically along boundary RG flows, and whose fixed point values agree with the boundary entropy. Another question is if $g$ can be defined directly in terms of an entropy. The conformal map of \cite{Calabrese:2009qy} identifies $g$ with a specific constant term in the entanglement entropy, after subtracting the logarithmically divergent area term. This subtraction obscures a possible monotonous behavior.

The goal of this work is to prove an entropic $g$-theorem, namely that there exists a $g$-function that decreases monotonically under boundary renormalization, and whose fixed point values agree with $g$ for BCFTs.\footnote{We would like to mention the previous related work \cite{Azeyanagi:2007qj}, where the authors attempted to prove the $g$-theorem using strong subadditivity. The holographic version of the theorem was established in \cite{Fujita:2011fp}.} We will accomplish this by identifying $g$ with a relative entropy; this is our main result and is presented in \S \ref{sec:relative}. In the remainder of the paper we initiate a broader program of using techniques from quantum information theory to study boundary RG flows. Specifically, in \S \ref{sec:mutual} we focus on the mutual information and how it measures correlations between the impurity and bulk degrees of freedom. In order to illustrate our general results, we introduce in \S \ref{sec:freeKondo} a new relativistic Kondo model, which has the nice feature of being Gaussian and yet it leads to a nontrivial boundary RG flow. Various aspects of quantum entanglement for this theory are analyzed in \S \ref{sec:application}.

\section{The entropic $g$-theorem from relative entropy}\label{sec:relative}

In this section we will study boundary RG flows using the relative entropy. The relative entropy provides a measure of statistical distance between the states of the system with different boundary conditions, and we will see that it is closely related to boundary entropy. Monotonicity of the relative entropy will be used to prove the $g$-theorem.

After reviewing boundary RG flows in \S \ref{subsec:boundaryRG}, in \S \ref{subsec:rel-review} we explain the connection between relative and boundary entropy. The relative entropy compares two density matrices: one corresponds to some arbitrary reference state (which in our case will be related to UV BCFT) and the other one is the density matrix for the system with relevant boundary flow. The simplest possiblity is to use reduced density matrices for intervals on the real line. We explore this in \S \ref{subsec:relative-real}, finding that the monotonicity properties of the relative entropy do not allow to prove a $g$-theorem. The reason is that the relative entropy distinguishes the different states too much, and this masks the decrease of $g$ under the RG. 

This suggests the correct path towards the $g$-theorem: vary the states in order to minimize the contribution from the modular Hamiltonian, while keeping fixed the entanglement entropy. This analysis is presented in \S \ref{subsec:proof}. We show that by working with states on the null boundary of the causal domain, the contribution from the modular Hamiltonian becomes a constant, and hence the impurity entropy is given explicitly as (minus) a relative entropy. We then use this result to prove the entropic $g$-theorem.

\subsection{Boundary RG flows}\label{subsec:boundaryRG}

Let us begin by briefly reviewing the class of RG flows that are studied in this work. The starting point is a CFT defined on $x_1>0$ with a boundary at $x_1=0$ which preserves half of the conformal symmetries -- a BCFT. This requires
\be
-i T_{01}(x_1=0)=T(x_1=0) - \bar T (x_1=0)=0\,.
\ee
A particular case is a CFT with a defect at $x_1=0$, which can be folded into a BCFT on $x_1 \ge 0$.\footnote{The reverse, unfolding a BCFT into a theory defined on the full line, is not possible in general. We thank E. Witten for pointing this out to us.}

In general the boundary may support localized degrees of freedom that will be coupled to the fields in the bulk theory. The UV theory, denoted by  BCFT$_{UV}$, is then perturbed by a set of relevant local operators at the boundary,
\be\label{eq:boundaryRG}
S= S_{BCFT_{UV}}+\int dx_0\, \lambda_i \phi_i(x_0)\,.
\ee
This perturbation can combine operators from the bulk (evaluated at $x_1=0$) and/or quantum-mechanical degrees of freedom from the impurity. The perturbation triggers a boundary RG flow; we assume that the flow ends at another boundary CFT, denoted by BCFT$_{IR}$.

The boundary perturbation preserves time-translation invariance and is local. In this way, bulk locality is preserved and operators at spatially separated points commute. This is needed for using the monotonicity of the relative entropy below.

The boundary entropy $\log g$ is defined as the term in the thermal entropy that is independent of the size of the system \cite{Affleck:1991tk},
\be
S= \frac{c\pi}{3 }\,\frac{L}{\beta}+\log g\,,
\ee
where $L$ is the size and $\beta$ the inverse temperature. At fixed points, $\log g$ can be computed as the overlap between the boundary state that implements the conformal boundary condition and the vacuum \cite{Cardy:1989ir}. For a boundary RG flow, Affleck and Ludwig conjectured that 
\be
\log g_{UV} > \log g_{IR}\,.
\ee 
Friedan and Konechny \cite{Friedan:2003yc} proved nonperturbatively that the boundary entropy decreases monotonically along the RG flow,
\be
\mu \frac{\partial \log g}{\partial \mu}\le 0\,,
\ee
where $\mu$ is the RG parameter. 
 It also decreases with temperature, since $g=g(\beta \mu)$ on dimensional grounds.

At a fixed point, the thermal entropy can be mapped to an entanglement entropy by a conformal transformation --see e.g. \cite{Calabrese:2009qy}. Concretely, The ground state entanglement entropy of an interval $x_1\in[0,r)$, with one end point attached to the boundary, is given by 
\be\label{eq:Sg1}
S(r) = \frac{c}{6}\,\log\,\frac{r}{\epsilon}+c_0 +\log g\,,
\ee
where $\epsilon$ is a UV cutoff, and $c_0$ is a constant contribution from the bulk that is independent of the boundary condition.

Therefore,  $\log g_{UV} > \log g_{IR}$ for the constant term on the entanglement entropy of this interval. 
 However, away from fixed points the entanglement entropy cannot be mapped to a thermal entropy, and it is not known whether $\log g(r)$ defined in (\ref{eq:Sg1}) decreases monotonically. We will prove that this is indeed the case.

\subsection{Boundary entropy from relative entropy}\label{subsec:rel-review}

The relative entropy between two density matrices $\rho_0$ and $\rho_1$ of a quantum system is defined as
\be\label{eq:def-relative}
S_{rel}(\rho_1 | \rho_0)=\tr(\rho_1 \log \rho_1)-\tr(\rho_1 \log \rho_0)\,.
\ee
In terms of the modular Hamiltonian for $\rho_0$, $\rho_0= e^{-\mathcal H}/tr(e^{-\mathcal H})$, it can be written as
\be\label{eq:relative2}
S_{rel}(\rho_1 | \rho_0)= \Delta \langle \mathcal H \rangle -\Delta S\,,
\ee
where $\Delta \langle \mathcal H \rangle= \tr \left((\rho_1-\rho_0)\mathcal H \right)$, and $\Delta S= S(\rho_1)-S(\rho_0)$ is the difference between the entanglement entropies of the density matrices.

Let us recall some basic features of the relative entropy.\footnote{We refer the reader to \cite{NC} for more details.}
For our purpose, the most relevant property of the relative entropy is that (for a fixed state)  it cannot increase when we restrict to a subsystem. In QFT the reduced density matrix $\rho_V$ is associated to a region $V$ and is obtained by tracing over the degrees of freedom in the complement $\bar V$. In this case, the relative entropy increases when we increase the size of the region. Some simple properties of the relative entropy are that $S_{rel}(\rho_1 | \rho_0)=0$ when the states are the same, and $S_{rel}(\rho_1 | \rho_0)=\infty$ if $\rho_0$ is pure and $\rho_1 \neq \rho_0$.

For the boundary RG flows of \S \ref{subsec:boundaryRG}, the reduced density matrix associated to an interval $x_1 \in [0, r)$ is obtained by tracing over the complement, and defines a $g$-function
\be\label{eq:gr}
S(r) = \frac{c}{6}\,\log\,\frac{r}{\epsilon}+c_0+\log g(r)\,.
\ee
This boundary entropy interpolates between $\log g_{UV}$ for $r \ll \Lambda^{-1}$ and $\log g_{IR}$ for $r \gg \Lambda^{-1}$. Here $\Lambda$ is the mass scale that characterizes the boundary RG flow. We want to show
\be
g'(r)\le 0\,,
\ee
and this would imply the entropic version of the $g$-theorem. Note that even if the theorem gives a monotonicity $g(0)\ge g(\infty)$ between fixed points, and coincides in this respect with the result \cite{Friedan:2003yc}, the interpolating function differs from their interpolating function. Indeed, as emphasized before, the boundary contribution in the thermal entropy does not map simply into the boundary contribution to the entanglement entropy when the theory is not conformal.

Let $\rho$ be the reduced density matrix on the spatial interval $[0, r)$. We want to compare $\rho$ with some appropriately chosen reference state $\rho_0$ in terms of the relative entropy. Since the boundary RG flow starts from a BCFT in the UV, we choose the reduced density matrix $\rho_0$ to be that of BCFT$_{UV}$. A crucial property that motivates this choice is that the modular Hamiltonian $\mc H$ for an interval including the origin in half space with a conformal boundary condition is local in the stress tensor, and has the same form as that of a CFT in an interval. This can be shown by a conformal mapping to a cylinder \cite{Calabrese:2009qy,jensen00}; see \cite{Cardy:2016fqc} for a recent discussion.\footnote{We thank J. Cardy for explanations on this point.}

Making this choice obtains
\be\label{eq:Srel-general}
S_{rel}(\rho|\rho_0)=-\log\,\frac{g(r)}{g(0)}+\tr\left((\rho-\rho_0)\mc H_{BCFT} \right)\,.
\ee
The first term comes from the difference in entanglement entropies between the theory with boundary RG flow ($\rho$) and the UV fixed point $\rho_0$; from (\ref{eq:gr}) this gives precisely the change in boundary entropy.
This gives the relation between the boundary entropy and relative entropy, and has the right sign to yield $g'(r)<0$ since $S_{rel}$ increases with $r$. The second term, however, could be an important obstruction to a $g$-theorem. It comes from the difference in expectation values of the modular Hamiltonian between the states with and without the relevant boundary perturbation. The rest of the section is devoted to analyzing this contribution. For the simplest setup of states defined on the real line, we will find that this term increases with $r$, masking the monotonicity of $g$. We will then improve our setup, showing how this term can be made to vanish by defining states on null lines.

\subsection{Relative entropy for states on the real line}\label{subsec:relative-real}

We have to understand the contribution of the modular Hamiltonian to (\ref{eq:Srel-general}). The simplest possibility is to work with states defined on the real $x_1$ line. In this case, the modular Hamiltonian for a CFT in half-space with a conformal boundary condition at $x_1=0$ is
\be
\mathcal H_{BCFT}(r)=2 \pi\int_0^r dx_1\, \frac{r^2-x_1^2}{2r} T_{00}(x_1)\label{h}\,. 
\ee 
This is the generator of a one parameter group of conformal symmetries that map the $x_1=0$ line in itself and keeps the end point of the interval $x_1=r,t=0$, fixed. These global symmetries of the CFT continue to be symmetries of the CFT with conformal boundary conditions. 

It is important that even in presence of a relevant perturbation on the boundary we must have $\langle T_{00}\rangle=0$ outside the $x_1=0$ line. This follows from tracelessness, conservation, and translation invariance in the time direction, that give
\bea
\langle T_{00}\rangle -\langle T_{11}\rangle&=&0\,, \nonumber \\
\partial_0 \langle T_{00}\rangle-\partial_1 \langle T_{10}\rangle=-\partial_1 \langle T_{10}\rangle&=&0\,,\\
\partial_0 \langle T_{01}\rangle-\partial_1 \langle T_{11}\rangle=-\partial_1 \langle T_{11}\rangle&=&0\nonumber\,.
\eea 
Hence $\langle T_{\mu\nu}\rangle $ is constant outside the boundary and has to vanish. 

Then $\langle T_{00}\rangle$ does not contribute to $\mathcal H_{BCFT}$ outside the boundary. If this is the whole contribution to $\Delta \langle \mathcal H_{BCFT}\rangle$ we would have from (\ref{eq:Srel-general}) that the monotonicity of the relative entropy implies the entropic $g$-theorem. In particular, $-g(r)$ would be given by the relative entropy between states with and without the boundary perturbations.

There is still an important aspect to understand: there might be a contribution to $\langle T_{00}\rangle$ localized at the boundary. On dimensional grounds, we expect for the variation of the expectation values with and without the relevant perturbation
\be
\Delta \langle T_{00}\rangle=\lambda^2 \epsilon^{1-2\Delta} \,\delta(x_1)+\ldots \label{consist}
\ee
where $\lambda$ is the relevant boundary coupling in (\ref{eq:boundaryRG}) with scaling dimension $[\lambda]=1-\Delta>0$, and $\epsilon$ is a distance cutoff. In other words, the boundary operator $\phi$ that deforms the theory in (\ref{eq:boundaryRG}) has dimension $\Delta$. Here we have done a perturbative expansion for small $\lambda$, so that $\rho$ and $\rho_0$ are very close to each other; the first perturbative contribution is generically of order $\lambda^2$. By a similar power-counting argument, more singular contact terms (proportional to $\lambda^2\epsilon^{2-2\Delta}\delta'(x_1)$ for example) would vanish in the continuum limit $\epsilon \to 0$.
From (\ref{h}) it is clear that any such localized contribution to $\langle T_{00}\rangle$ will produce a contribution to $\Delta \langle \mathcal H_{BCFT}\rangle$ which is increasing linearly with $r$, spoiling a proof of the $g$-theorem. In the free Kondo model of \S \ref{sec:freeKondo} we will see that this is indeed the case.

This linear dependence in $r$ implies that the relative entropy distinguishes too much the states with and without the impurity on the real line. It is clear that in order to be able to use the relative entropy to capture the RG flow of $g(r)$ we need to choose states that minimize $\Delta \langle \mc H_{BCFT} \rangle$. This is the problem to which we turn next.

\subsection{Proof of the entropic $g$-theorem}\label{subsec:proof}

In order to use the monotonicity of the relative entropy to prove the $g$-theorem, we need to minimize the contribution from the modular Hamiltonian. The basic idea is that in a unitary theory the entanglement entropy is the same on any spatial surface that has the same causal domain of dependence. 
This evident is in the Heisenberg representation, where the state is fixed and local operators depend on spacetime. Local operators written in a given Cauchy surface can in principle be written in any other Cauchy surface using causal equations of motion. Then, the full operator algebra written in any Cauchy surface will be the same, and as the state is fixed, the entropy will remain invariant. 

The relative entropy for two states in a fixed theory is also independent of Cauchy surface. 
However, in the present case, as the vacuum states of the theory with or without relevant boundary perturbation have different evolution operators, choosing a different surface corresponds to changing the states by different unitary operators in each case. In the Heisenberg representation of the BCFT the conformal vacuum will not change, but the fundamental state of the theory with the relevant perturbation will evolve with an additional insertion placed on $x_1=0$. As a consequence $\Delta\langle  \mc H_{BCFT} \rangle$ will now depend on the choice of surface. Therefore, we need to vary the Cauchy surface until we eliminate the large increasing $\Delta \langle \mc H_{BCFT} \rangle$ term in the relative entropy.

\begin{figure}[h!]
\begin{center}  
\includegraphics[scale=0.55]{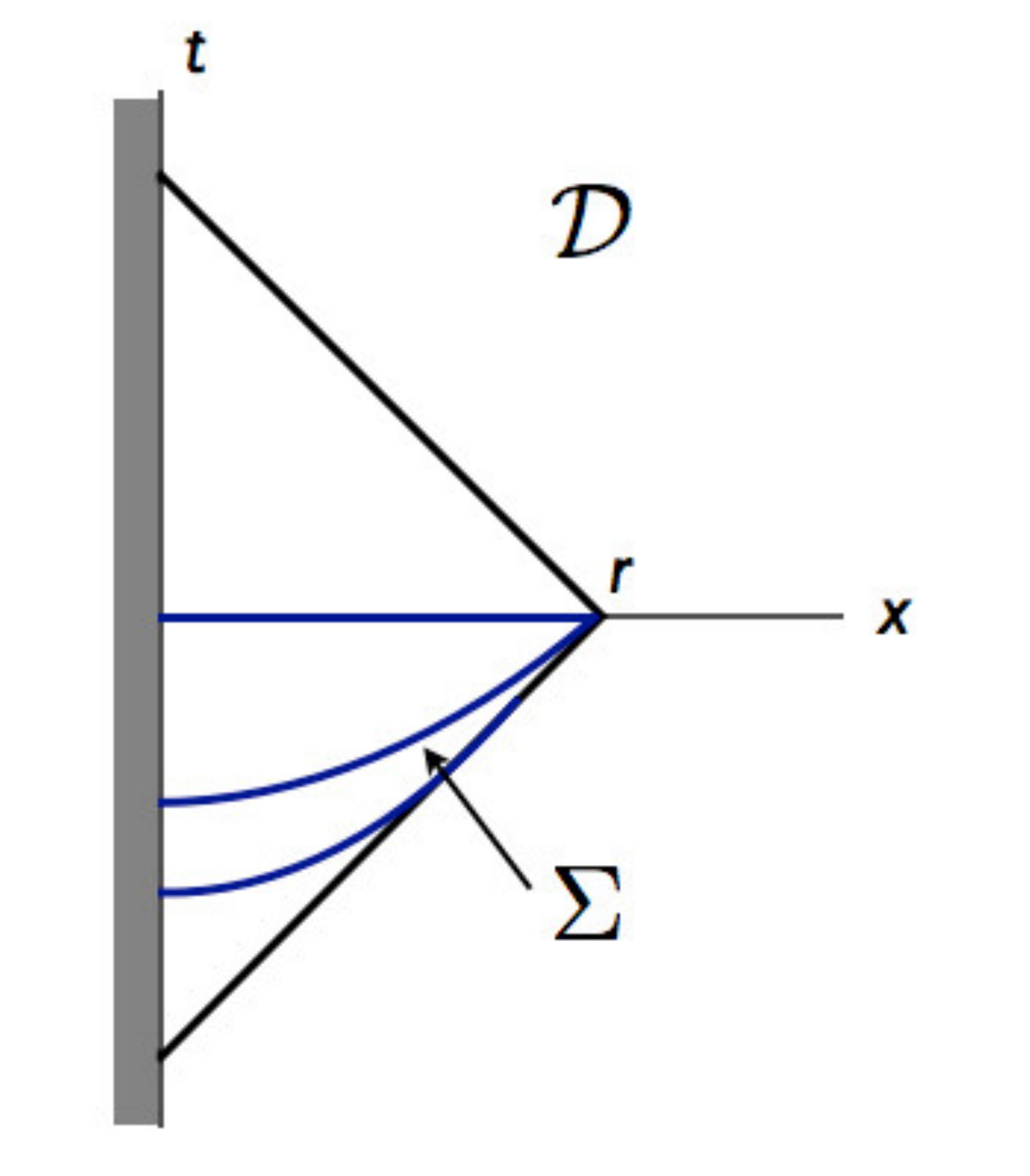}
\captionsetup{width=0.9\textwidth}
\caption{Different Cauchy surfaces $\Sigma$ with the same causal domain of dependence $\mc D$ give the same entanglement entropy $S(r)$.}
\label{fig:causal1}
\end{center}  
\end{figure}

This approach is illustrated in Figure \ref{fig:causal1}. We want to determine the entanglement entropy $S(r)$ for a spatial interval $x_1 \in [0,r)$. This interval defines a causal domain of dependence $\mc D$, and because of unitarity $S(r)$ is the same for any other Cauchy surface with the same $\mc D$. This applies for both states, since evolution is unitary inside $\mc D$ independently of the local term in the Hamiltonian at $x_1=0$. Hence, $\Delta S$ is independent of the chosen surface $\Sigma$. 

We want to make $\rho$ as similar as possible to $\rho_0$ in order to minimize the contribution $\Delta \langle \mc H_{BCFT} \rangle$. 
The modular Hamiltonian of the BCFT vacuum is proportional to the generator of conformal transformations that keep the interval fixed. Using the Heisenberg representation corresponding to the BCFT evolution, it can be written on any Cauchy surface $\Sigma$ as a flux of a conserved current
\be
{\cal H}_{BCFT}=\int_{\Sigma} ds \, \eta^\mu T_{\mu\nu} \xi^\nu\,,
\ee
where $\eta$ is the unit vector normal to the surface and
\be
\xi^\mu\equiv \frac{2\pi}{2r}(r^2-(x^0)^2-(x^1)^2\,,\,-2 x^0 x^1 )\,.
\ee 
We stress again the important point that this current is generally not conserved in the theory with boundary RG flow, leading to changes in the expectation values of the modular Hamiltonian for different surfaces. 

Since the expectation values of the stress tensor vanish everywhere except at the impurity we need to choose a surface where the coefficient of $T_{\mu\nu}$ in the modular Hamiltonian vanishes on the line $x^1=0$.
We accomplish this by working with a state on the null boundary of the causal development; see Figure \ref{fig:causal1}.
In null coordinates $x^\pm=x^0\pm x^1$ this writes  
\be
\mc H_{BCFT}=2\pi \int_{-r}^r\,dx^+\,\frac{r^2-x^{+\,2}}{2r} T_{++}(x^+)\,.
\label{nunu}
\ee
By locality, the defect at $x^+=x^-=-r$ can contribute a contact term of the form
\be\label{eq:Tpp}
\langle T_{++} \rangle\sim  \delta(x^++r)
\ee
and similarly for the $T_{--}$ component. This effect gives a vanishing contribution in (\ref{nunu}). This should be contrasted with the situation on the real line, where a delta function $\langle T_{00} \rangle \sim \delta(x_1)$ already contributes a linear term in $r$ to the modular Hamiltonian (\ref{h}).

We conclude that, by working with a state on the null segment, the contribution from $\Delta \langle \mc H \rangle$ vanishes 
\be\label{eq:gteo}
S_{rel}(\rho|\rho_0)=-\log\,\frac{g(r)}{g(0)}\,.
\ee
The change in the boundary entropy is then identified as a relative entropy. Note that with the relative entropy we can measure changes in the boundary entropy, and not the boundary entropy itself.

In physical terms, the reason that relative entropy is much smaller in the null surface than in the spatial one is that in this last case we are placing the impurity at the origin of the interval where the vacuum of the BCFT has an effective low temperature $\sim r^{-1}$ as can be read off from the coefficient of $T_{00}$ in (\ref{h}). As a result the two states are highly distinguishable, having a large relative entropy. In contrast, the extreme point of the null Cauchy surface (corresponding to $x^+=-r$) is a point of an effective high temperature, as seen from the fact that the coefficients of $T_{++}$ vanish there in (\ref{nunu}). Hence distinguishability is strongly reduced, and will be driven by the change of correlations outside the impurity, which will be reflected in the change of entanglement entropies.

\begin{figure}[hb!]
\begin{center}  
\includegraphics[scale=0.4]{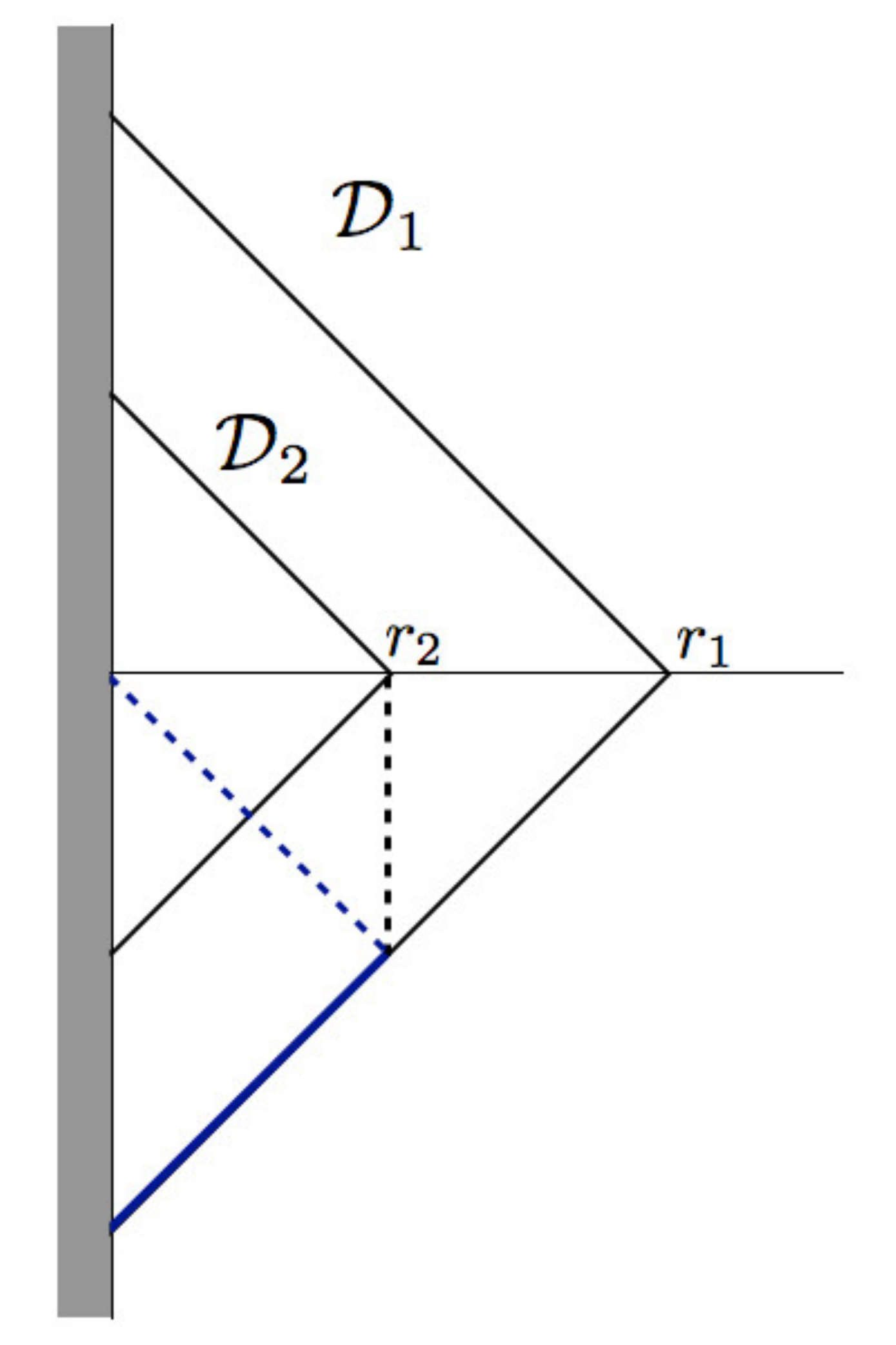}
\captionsetup{width=0.9\textwidth}
\caption{Using time-translation invariance, the smaller causal domain of dependence $\mc D_2$ is translated so that its past null boundary overlaps with that of $\mathcal D_1$. We then have the same state specified on the null boundary, and varying $r$ gives an increasing relative entropy.}
\label{fig:causal2}
\end{center}  
\end{figure}

Finally, in order to use the monotonicity of the relative entropy, we need to vary $r$ but using the same states defined on the null line. This is implemented as explained in Figure \ref{fig:causal2}.
The monotonicity of the relative entropy gives
\be
g'(r)<0\,.
\ee
This completes our proof of the entropic $g$-theorem. The relative entropy defines a monotonic $g$-function, and the total change between the UV and IR boundary CFTs is 
\be
S_{rel}(\infty)-S_{rel}(0)=\log (g_{UV}/g_{IR}) >0\,.
\ee
This formula is independent of contact terms and establishes a universal relation between the change in relative entropy and the total running of the boundary entropy. 

In this proof of the $g$-theorem we have compared the density matrix $\rho$ along the RG flow to the state $\rho_0$ of the UV fixed point. This was used, in particular, to constrain the form of the contact term divergences in (\ref{eq:Tpp}). While in our context this is the most natural choice for $\rho_0$, one may wonder what happens if $\rho_0$ is some other reference state. One possibility along these lines is to use the IR BCFT as the reference. For large enough $r$, $\rho$ approaches $\rho_0$ on the null line, and the contributions to $\Delta \langle T_{\mu\nu}\rangle$ are determined by the leading irrelevant operator that controls the flow towards the IR fixed point. This flow does not have a well-defined UV limit, and hence other divergences besides (\ref{eq:Tpp}) are allowed. In particular, at least a contact term proportional to $\delta'(x^++r)$ is required in order to ensure the positivity of the relative entropy. Unlike the choice of $\rho_0$ as the state of the UV fixed point, the contribution from $\Delta \langle \mc H \rangle$ will then generically be nonzero and this will obscure the monotonic behavior of $\log g(r)$. Similar remarks apply to other choices of $\rho_0$.

\section{Mutual information in quantum impurity systems}\label{sec:mutual}

In the previous section we related the boundary entropy to the relative entropy, and proved an entropic $g$-theorem. We now explore another measure from quantum information theory, the mutual information. In QFT, the mutual information between regions $A$ and $B$ is given in terms of the entanglement entropy by
\be
I(A,B)=S(A)+S(B)-S(A\cup B)\,.
\ee
Mutual information is always positive and increasing with region size. It has the interpretation of shared information (classical and quantum) between the two regions. 

There are two important motivations for considering the mutual information in the context of quantum impurity systems. The first motivation is that it provides a measure of the correlations in the system. In more detail, it is a universal upper bound on correlations \cite{vers}
\be\label{eq:mutual-bound}
I(A, B) \ge \frac{\left(\langle O_A O_B \rangle-\langle O_A\rangle \langle O_B\rangle \right)^2}{2 \left\|O_A\right\|^2 \left\|O_B\right\|^2} 
\ee
for bounded operators $O_A$ and $O_B$ that act on $A$ and $B$ respectively.  The second reason is the connection with the boundary entropy $\log g$. 

Our proposal is to study the dynamics of quantum impurity systems in terms of the mutual information between the impurity (subsystem A above) and an interval of size $r$ in the bulk (subsystem B). We first discuss in \S \ref{subsec:Icorrelations} why and how this mutual information captures correlations between the impurity and bulk degrees of freedom. We then consider the relation between boundary entropy and mutual information. This is illustrated in \S \ref{subsec:valence} in terms of a toy model of a lattice of spins with bipartite entanglement, where $I(A, B)$ and $\log g$ are related explicitly. Section \ref{subsec:general-proposal} gives a more general discussion of mutual information in the presence of impurities.

\subsection{Mutual information and correlations}\label{subsec:Icorrelations}

Let us analyze the connection between mutual information and correlations. For this, we consider first a continuum QFT without impurity and argue that generically the mutual information will vanish when the size of $A$ above goes to zero. We then add the impurity, contained in $A$, and discuss how the new correlations between this quantum-mechanical system and the bulk will manifest themselves in a nontrivial mutual information.

In QFT, the mutual information $I(A,B)$ between two regions will generally go to zero if $A$ is made to shrink to a point $y$ while keeping $B$ constant. The reason is that all fixed operators in $A$ whose correlations with operators in $B$ are non zero will eventually drop out from the algebra of $A$. In this sense we recall that in order to construct a well defined operator in the Hilbert space localized in $A$, we have to smear the field operators, $\phi_A=\int dx\, \alpha(x) \phi(x)$ with a test function $\alpha(x)$ with support in $A$. Thus, even if $\phi(y)$ is always present in $A$ as we take the limit $A\rightarrow y$  this is not a bounded operator living in the algebra of local operators in $A$. 

Let us illustrate how this happens for a CFT in $d=2$. Take two intervals $A$ and $B$ of size $a$ and $b$ respectively, separated by a distance $c$. Mutual information is conformal invariant and will be a function $I(\eta)$ of the cross ratio 
\be
\eta=\frac{a b}{(a+c)(b+c)}\,.
\ee   
Then, as we make $a\rightarrow 0$ keeping $b,c$ constant we have $\eta\rightarrow 0$. To evaluate this limit we can think in another configuration with the same cross ratio, for example taking $a^\prime=b^\prime=1$, $c^\prime=1/\sqrt{\eta}-1$, which diverges as $\eta^{-1/2}$ for small $\eta$. This is two unit intervals separated by a large distance. In this case, the mutual information will vanish as 
\be\label{eq:mutual-asymp}
I(\eta)\sim \eta^{2\Delta}\sim a^{2\Delta}\,,
\ee
where $\Delta$ is the minimum of the scaling dimensions of the theory \cite{Cardy}.\footnote{The case $\Delta=0$ corresponds to the free massless scalar, that is not a well defined model --in particular the zero mode makes the mutual information for any regions infrared divergent.} 

This is consistent with mutual information being an upper bound on correlations. 
 If we find any bounded operators, normalized to norm one, with non zero connected correlator, the mutual information cannot be zero. If for $a$ going to zero the mutual information goes to zero it must be that all correlators (for normalized operators) go to zero. Let us try with a smeared field $\phi_\alpha=\int dx\, \phi(x) \alpha(x)$, constructed with a $\phi$ of scaling dimension $\Delta$. $\phi_\alpha$ is not generally bounded, and this would unfairly give zero to the right hand side of (\ref{eq:mutual-bound}) even for fixed finite size intervals. We can circumvent this problem by doing a spectral decomposition of the operator and using an operator $\tilde{\phi}_\alpha$ that is $\phi_\alpha$ up to some cutoff in the spectral decomposition. We choose this cutoff such that the correlators of $\phi_\alpha$ with itself and $O_B$ at the separations of interest are well reproduced by $\tilde{\phi}_\alpha$. We have that 
\be
\langle 0|\tilde{\phi}_\alpha \tilde{\phi}_\alpha|0\rangle\le \left\|\tilde{\phi}_\alpha\right\|^2
\ee
because $\left\|\tilde{\phi}_\alpha\right\|^2$ is the supremum of the expectation value of $\tilde{\phi}_\alpha \tilde{\phi}_\alpha$ for all unit vectors in the Hilbert space. Then, the right hand side of (\ref{eq:mutual-bound})  for this operator is smaller than
\be
\frac{\left( \int_A dx\, \alpha(x)\,\langle \phi(x) O_B \rangle \right)^2}{2 \left\|O_B\right\|^2\int_A dx\, dy\, \frac{\alpha(x)\alpha(y)}{|x-y|^{2\Delta}} }\sim a^{2 \Delta}\,
\ee 
which is compatible with (\ref{eq:mutual-asymp}). 

We see that the fact that correlators of fields diverge at short distances is important in this argument. In fact, if that were not the case, the field at a single point itself would be a well defined operator in Hilbert space, and mutual information between this point and another system could have a non zero value. While this is not the case of continuum QFT, this is clearly the case of an ordinary quantum mechanical degree of freedom (in $0+1$ dimensions) since all field operators $\phi(t)$ are operators in the Hilbert space (as opposed to operator valued distributions) and have finite correlators $\langle \phi(t)\phi(t^\prime)\rangle$ for $t^\prime\rightarrow t$. 

Systems with impurities fall precisely in this category.
Then, the mutual information of a region of the QFT with an interval $[0,a)$ containing the quantum mechanical degrees of freedom of the boundary theory can have a non trivial limit as $a\rightarrow 0$. Of course, in systems with no degrees of freedom living at the boundary, the mutual information wouldn't yield a useful measure, by our arguments above. However, in order to produce nontrivial boundary RG flows, we generically expect that such degrees of freedom will be needed, and hence the mutual information would provide a useful characterization of the dynamics. One way to diagnose this is to determine if the bulk is pure along the RG; if it is not pure, then purifying it with a system $A$ we regain the possibility of obtaining a nontrivial mutual information. A simple example of this situation is illustrated in \S \ref{sec:freeKondo}. 

In summary, our proposal is to look at the mutual information 
\be
I([0,\epsilon],[\epsilon^\prime,r])
\ee
where $\epsilon$ is a short cutoff, and can in fact be set to a microscopic distance or just consider the boundary degrees of freedom. $\epsilon^\prime$ is another microscopic distance greater than $\epsilon$. As we increase $r$ this quantity will increase with $r$. Possible short distance correlations across $[\epsilon,\epsilon^\prime]$ will give an overall constant term to the mutual information which will not change with $r$. This can be set to zero just using microscopic distances, or a large ratio $\epsilon^\prime/\epsilon$. 

\subsection{Impurity valence bond model}\label{subsec:valence}

To motivate our proposal, and in order to understand how this works out, we consider a simple spin system with bipartite entanglement. For this case the impurity entropy is captured directly by the mutual information and is monotonic along boundary RG flows.

A lattice model that is equivalent to the Kondo model in the continuum is a 1d spin system with nearest and second-nearest neighbor hopping terms, and an impurity in the first site \cite{affleck1}. The Hamiltonian is
\be
H= J' (\vec S_1 \cdot \vec S_2+J_2 \vec S_1 \cdot \vec S_3)+\sum_{j=2}^{N-1} \vec S_j \cdot \vec S_{j+1}+J_2\sum_{j=2}^{N-2} \vec S_j \cdot \vec S_{j+2}\,.
\ee
The impurity corresponds to the first site with $J' \neq 1$, and all the spins $s=1/2$. 

The impurity entanglement entropy for a subsystem $R$ with sites $j=1, \ldots, r$, which contains the impurity at one end, is defined on the lattice as
\be\label{eq:gdef1}
\log\,g(r)=S(r, J', N)-S(r-1, J'=1, N-1) \,.
\ee
where $S(r, J', N)$ is the entanglement entropy obtained by tracing out over sites $r+1, \ldots, N$, and $S(r-1, J'=1, N-1)$ is the same quantity but in a system with no impurity --this is accomplished by setting $J'=1$ and deleting one site.

Let us instead consider a different quantity: the mutual information between subsystem A --the impurity at site $j=1$-- and subsystem B comprised of sites $j=2, \ldots, r$. It is given in terms of the entanglement entropy (EE) by
\be\label{eq:mutual}
I(A, B)= S(A)+ S(B)-S(A \cup B)\,.
\ee

We model the entanglement in the theory in terms of an ``impurity valence bond'', as in \cite{affleck1}. This is the bond that connects the impurity and the other spin in the lattice with which it forms a singlet. Let's denote this site by $k$. This provides a simple intuition for the impurity entanglement entropy: if the interval $R$ (which contains sites $j=1, \ldots, r$) cuts the bond, this gives a $\log 2$ contribution to the EE, while the impurity entanglement vanishes if the bond is inside $R$. Then if $1-p$ is the probability that the impurity valence bond is cut by $R$, namely $k \not \in R$ with probability $1-p$, we may write (\ref{eq:gdef1}) as
\be\label{eq:EEbond}
\log \,g(r) = (1-p(r))\,\log 2+p \left(S_\text{no imp}(r-2) -S_\text{no imp}(r-1)\right)\,,
\ee
where $S_\text{no imp}$ is the EE with $J'=1$. We also assume that $N$ is sufficiently large, such that the difference in entanglement entropies with $N$ and $N-1$ is negligible. This is then a simplified picture in terms of a probabilistic distribution of bipartite entanglement. In particular, in the continuum limit we expect 
\be\label{eq:gp}
\log g(r) = (1-p(r)) \log 2\,.
\ee

We now evaluate the mutual information (\ref{eq:mutual}) in terms of the impurity valence bond. First,
\be
S(A) = \log 2 = \log g(r=0)
\ee
is the total impurity entanglement, or the $g$-function in the UV. The $S(B)$ term gets a contribution $p \log 2$ from entanglement with the impurity, plus the entanglement with the rest of the system, i.e.
\be
S(B) = p \left(\log 2+S_\text{no imp}(r-2)\right)+(1-p)S_\text{no imp}(r-1)\,.
\ee
Here, with probability $p$ the impurity spin is entangled with one of the spins in B, and then the rest of the spins in B ($r-2$ of them) is entangled with the rest of the system as if there were no impurity. With probability $1-p$ the impurity spin is entangled with one of the spins outside B, and hence the EE for B, which has $r-1$ sites, is the EE with a system of $N-1$ spins and no impurity. In the continuum limit, the difference between the intervals of size $r-1$ and $r-2$ will be negligible, and hence
\be
S(B) = p(r) \log 2+S_\text{no imp}(r)\,.
\ee
Similarly, for $S(A\cup B)$ with probability $p$ one spin in B is entangled with A and hence the entanglement with the rest is $S_\text{no imp}(r-2,N)$, while with probability $1-p$ the valence bond is outside $A \cup B$ and the entanglement with the rest of the system is $S_\text{no imp}(r-1,N-1)$. Taking the continuum limit obtains
\be\label{eq:Sunion}
S(A\cup B)=(1-p(r)) \log 2+S_\text{no imp}(r)\,.
\ee
Notice that $S(A \cup B)$ contains $\log g(r)$, while the impurity contribution in $S(B)$ is $\log 2- \log g(r)$. This simplification is a consequence of bipartite entanglement and will not occur in the multipartite case.

Putting these contributions together and writing $p(r)$ in terms of $\log g(r)$ obtains, in the continuum,
\be
I(A, B)= 2 \,\log\frac{g(0)}{g(r)}\,.
\ee
Since the mutual information is non-increasing under discarding parts of the system, it follows that
\be
\frac{dg}{dr} \le 0\,.
\ee
In other words, the entropic $g$-function defined in terms of mutual information decreases monotonically under RG flows in this simplified picture of bipartite entanglement.

While this model is of limited applicability, it serves to illustrate the connection between mutual information and boundary entropy. We will next study more general systems allowing for multipartite entanglement.

\subsection{General analysis}\label{subsec:general-proposal}

We learned from the previous simplified model that the running of the constant term in the entropy is due to entanglement with an impurity. This impurity has the effect of changing boundary conditions from a preexisting one in the UV to a different one in the IR. The full system formed by $A$, and the line $x_1>0$  is pure. As $g(0)$ is the impurity entropy in the UV, $g(\infty)$ measures the residual entropy in the impurity that has not been neutralized by entanglement with the field as we move to larger $r$. 

The important question that remains is how to generalize this argument to include multipartite entanglement. 
In the mutual information we will have, in general
\be
S(A \cup B) = \log\,g(r)+S_\text{no imp}(B)\,.
\ee
A new quantity, $\tilde g$, appears for the EE of B in the presence of the impurity:
\be
S(B) = \log\,\tilde g(r) +S_\text{no imp}(B)\,,
\ee
and then
\be
\label{Irlogg}
I(A,B) = I(r)=S(A) +\log\,\tilde g(r)- \log\,g(r)\,.
\ee
This must be an increasing function. We know $g(r)$ goes from the conformal value $g(0)$ in the UV to the one $g(\infty)$ in the IR. 

The function $\tilde{g}(r)$ is determined by the entropy of $B$. The value of $\t g$ at the fixed points can be determined as follows.
Mutual information between the impurity and a small $B$ will be zero since correlations of the impurity are with regions further in the bulk. Hence $I(0)=0$ and
\be
S(A)=\log\, g(0)-\log\,\tilde{g}(0)\,. 
\ee
For large $r$ the entanglement of the impurity with local degrees of freedom at distance $r$ vanishes and the mutual information stops increasing. Hence $I(\infty)=2 S(A)$, 
\be
S(A)=\log \,\tilde{g}(\infty)-\log\, g(\infty)\,.
\ee
However, for finite nonzero $r$ in general there will be no simple relation between $g(r)$ and $\tilde g(r)$. In the previous model based on a probabilistic distribution of bipartite entanglement, $\t g(r)= g(0)-g(r)$, but we do not expect this to hold in the presence of multipartite entanglement. We analyze this for the free Kondo model in \S \ref{sec:freeKondo}.

The appearance of the new function $\t g(r)$ does not allow us to establish the monotonicity of the boundary entropy in terms of the mutual information --only the combination $\log \t g(r)-\log g(r)$ has to increase. In fact, this problem is related to what we found for the relative entropy in \S \ref{sec:relative}. To see this, we recall that the mutual information is a specific relative entropy,
\be
I(A,B)= S_{rel}(\rho_{AB}| \rho_A \otimes \rho_B)\,,
\ee
where $\rho_A =\tr_B \,\rho_{AB}, \rho_B =\tr_A \,\rho_{AB}$. We expect this quantity is dependent of the Cauchy surface. In fact, while $g(r)$ is related to the full entropy of the field coupled to the impurity, and cannot change with the surface because the evolution is unitary and causal for the full system, $\t g(r)$  does change with the surface --unitary evolution followed by partial tracing over the impurity does not keep the entropy constant. 
In fact, in \S \ref{sec:freeKondo} we will find that for a specific simple model the reference state on the null interval is that of the UV fixed point, and the mutual information is then the same as the relative entropy of \S \ref{subsec:proof}. In this case, $\log \t g(r)=0$ for all $r$, and
\be
I(r) = -\log\,g(r)+\text{const}\,.
\ee

The preceding argument exhibits the state dependence of $\t g(r)$, while $g(r)$ comes from the EE on the complete system and hence is surface-independent. Nevertheless, it would be interesting to understand in more detail the relation between $\t g(r)$ and multipartite entanglement; $\log g(r) + \log \t g(r)$ might give useful information on ``Kondo clouds'' \cite{cloud}.

\section{A free Kondo model}\label{sec:freeKondo}

Our task in this work has been to apply quantum information methods to the study of boundary RG flows in impurity systems, establishing the entropic $g$-theorem. In the remaining of the paper, we present a simple tractable model where we can illustrate our results. The Kondo model (see e.g. \cite{Affleck:1995ge, Ludwig:1994dy} for nice reviews) would be the ideal example for this, but this model is interacting; computing quantum information quantities requires then more advanced numerical tools which would go beyond the scope of our approach.\footnote{For a recent review of entanglement entropy in interacting impurity systems see \cite{affleck1}. It would be interesting to calculate mutual information and relative entropy in these systems using DMRG.}

Instead, in this section we construct a Gaussian model which reproduces the main feature of the Kondo model, namely the flow between `$+$' and `$-$' boundary conditions for the bulk fermions. The model is relativistic, though one may also consider a nonrelativistic version, closer to the Kondo system; this is described in Appendix \ref{app:nonrel}. Analytic and numeric calculations of quantum entanglement will be presented in \S \ref{sec:application}.

\subsection{The model}\label{subsection:freemodel}

Consider a two-dimensional Dirac fermion living in the half-space $x_1\ge 0$, interacting with a fermionic Majorana impurity at $x_1=0$ --a quantum mechanics degree of freedom:
\be\label{eq:model1}
S= \int_{-\infty}^\infty dx_0 \int_0^\infty dx_1\, \left(-i \bar \psi \gamma^\mu\partial_\mu\psi+\frac{i}{2}\delta(x_1) \left[\bar\chi \gamma^0\partial_0 \chi+m^{1/2}(\bar \psi \chi-\bar\chi \psi)\right] \right)\,.
\ee
The scaling dimensions are $[\psi]=1/2$, $[\chi]=0$, and hence $[m]=1$ and we have a relevant boundary perturbation. We emphasize that $\chi$ is a quantum mechanics degree of freedom and as such it scales differently than the bulk fermion. 

To understand the effects of the perturbation, we write the action in components, using the representation
\be
\gamma^0= \left(\begin{matrix}0 & 1\\ -1 & 0\end{matrix}\right)\;,\;\gamma^1= \left(\begin{matrix}0 & 1\\ 1 & 0\end{matrix}\right)\;,\;\psi= \left(\begin{matrix}\psi_+^*\\\psi_-\end{matrix}\right)\;,\;\chi= \left(\begin{matrix}\eta\\\eta^*\end{matrix}\right)\,.
 \ee
We work in signature $(-+)$. Note that $\gamma^5=\gamma^0 \gamma^1=\sigma_z$, and hence $\psi_\pm$ are the two chiralities in this basis. For later convenience, we have defined the left-moving component of $\psi$ as $\psi_+^*$. The resulting action is
\bea
S&=& \int_{-\infty}^{\infty} dx_0 \int_0^\infty dx_1\, \left(i \psi_+(\partial_0-\partial_1)\psi_+^*+i \psi_-^*(\partial_0+\partial_1)\psi_-\right.\nonumber\\
&&\left.+\delta(x_1) \left[i \eta^* \partial_0 \eta-\frac{i}{2}m^{1/2} \eta^*(\psi_++\psi_-)+c.c.\right]\right)\,.
\eea
The action is invariant under charge conjugation, as reviewed in Appendix \ref{app:conditions}.

In the UV, the boundary mass term is negligible compared to the boundary kinetic term, and hence we have a free quantum-mechanical fermion $\chi$, decoupled from the bulk system. Since the bulk lives in the half space, we need to impose a boundary condition that ensures the vanishing of the boundary term in the action variation. We choose,
\be\label{eq:plus}
\psi_+(x_0,0)= \psi_-(x_0,0)\,,
\ee
consistently with the charge-conjugation symmetry of the theory. This choice is also motivated by what happens in the interacting single-channel Kondo model; there the two chiralities come from the two points of the Fermi surface (in the radial problem), and they obey (\ref{eq:plus}) in the UV. We comment on more general boundary conditions in Appendix \ref{app:conditions}.

The interaction with the quantum-mechanics degree of freedom $\eta$ will induce a boundary RG flow in the form of a momentum-dependent reflection factor connecting the left and right moving bulk fermions. We will analyze this RG flow shortly. In the deep infrared, the boundary behavior simplifies: we may ignore the kinetic term of the impurity, treating it as a Lagrange multiplier that imposes
\be
\psi_+(x_0,0)= -\psi_-(x_0,0)\,.
\ee
Our explicit analysis below will verify this. Therefore, the free Kondo model gives an RG flow between the `$+$' and `$-$' boundary conditions. The same happens in fact in the interacting single-channel Kondo model. Our free model has the nice property of being completely solvable, and we will determine the RG flow --and various quantities from quantum information theory-- explicitly.

It is also possible to understand the dynamics of the impurity by integrating out the bulk fermions. This gives rise to an effective action at the boundary,
\bea\label{eq:Seffeta}
S_{eff}&=& \int dx_0\,\eta^* \partial_0 \eta\\
&+& \frac{m}{8}\int dx_0 dx_0'\,\left( \eta(x_0)\,G_+(x_0-x_0',0) \eta(x_0')+ \eta^*(x_0)\,G_-(x_0-x_0',0) \eta^*(x_0')\right)\,, \nonumber
\eea
where $G_\pm= (i \partial_\pm)^{-1}$ are the chiral propagators. At early times (UV), the tree level term dominates and $\dim(\eta)=0$; its propagator is just a constant. At late times (IR), the dynamics is dominated by the effective contribution from the bulk fermions. Since $G_\pm(t,0) \propto 1/t$ (the Fourier transform of $\Theta(p_0)$, we obtain a conformal quantum mechanics with $\dim(\eta)=1/2$ and
\be\label{eq:etaprop}
\langle \eta(x_0) \eta(x_0')\rangle \propto \frac{1}{m(x_0-x_0')}\,.
\ee 
This is to be contrasted with the interacting single-channel Kondo problem, where the impurity is confined in the IR.

\subsection{Lattice version}\label{subsection:lattice}

In order to compute entanglement entropies, let us now put the previous theory on a lattice. Due to fermion doubling, it is sufficient to consider a one-component bulk fermion interacting with the impurity:
\be\label{eq:lattice1}
L_{lattice}= a\sum_{j=0}^\infty \left(i \psi_j^* \partial_0 \psi_j-\frac{i}{2a}(\psi_j^* \psi_{j+1}-\psi_{j+1}^* \psi_j)\right)+i \eta^* \partial_0 \eta-\frac{i}{2} m^{1/2}(\eta^* \psi_0+c.c.) \,.
\ee
The hopping term comes from discretizing the symmetrized derivative operator $\frac{i}{2}\psi^* \partial_x \psi$. Setting the lattice spacing $a=1$, the quadratic kernel becomes
\be
M=\left(\begin{matrix}
0 & \frac{i}{2} m^{1/2} & 0 & 0 &\ldots \\
-\frac{i}{2} m^{1/2} & 0 & \frac{i}{2} & 0 &\ldots \\
0 & - \frac{i}{2} & 0 & \frac{i}{2} & \ldots \\
 0 & 0 &- \frac{i}{2} & 0 &  \ldots  \\
 \vdots & \vdots & \vdots &  &\ddots
\end{matrix}\right)\,.
\ee
The first site corresponds to the impurity. 
Note that for $m=1$ we have a lattice with one more site and no impurity --the quantum mechanics degree of freedom becomes the same as one of the discretized bulk modes. On the other hand, for $m=0$ the impurity decouples from the lattice system.

Let us first study the spectrum of this theory, in order to determine its relation with the previous continuum model. Thinking of $\eta$ as associated to an extra lattice point at $j=-1$, we construct $\Psi_j=(\eta, \psi_{j\ge 0})$ and look for eigenvectors
\be\label{eq:Meq}
M_{ij} \Psi_j(k) = E(k) \Psi_i(k)\,.
\ee
Looking at the sites $j \ge 1$, the solutions are combinations of incoming and outgoing waves,
\be\label{eq:eigenvectors}
\Psi_j(k) = a_k e^{i k j}+b_k (-1)^j e^{-i j k}
\ee
with eigenvalues
\be
E(k) = -\sin k\,.
\ee
The boundary condition chooses a specific combination of the momentum $k$ and $\pi-k$, with degenerate energies. Hence, the different eigenvectors are with $-\pi/2 <k<\pi/2$, $E=-\sin(k)$.
Since we have a different degree of freedom at $j=-1$, $\Psi_{-1}$ is not of this form. Evaluating (\ref{eq:Meq}) for $i=-1$ gives
\be\label{eq:eta1}
\eta= \Psi_{-1}=-\frac{i}{2}\,\frac{m^{1/2}}{\sin k} \psi_0\,.
\ee
On the other hand $M_{0j}\Psi_j =-\sin k\, \psi_0$ determines
\be
\frac{i}{2}\psi_1= \left(\frac{m}{4 \sin k}-\sin k \right) \psi_0\;\Rightarrow\;\frac{b_k}{a_k}=R(k)=-\frac{1-m-e^{-2i k}}{1-m-e^{2i k}}\,.
\ee
This gives the reflection coefficient at the wall, relating the left and right moving modes. In the UV, $m \to 0$ and
\be
R(k)= e^{-2 i k}\,.
\ee
This approaches $1$ in the continuum limit. On the other hand, in the IR $m \to \infty $ and
\be
R(k)=-1\,.
\ee

In terms of the annihilation operators $d_k$ of the modes of definite energy $k$ we have
\be
\psi_j=\sum_k  \psi_j(k)  d_k \,,
\ee
where $\psi_j(k)$ is given by (\ref{eq:eigenvectors}). The knowledge of the spectrum allow us to obtain the exact correlations functions for the infinite lattice, without the need of imposing an IR cutoff. This is important to compute the entropies which for a Gaussian state depend only on the two-point correlators on the region.

As we see from (\ref{eq:eigenvectors}), the lattice fermion $\Psi_j$ contains both L and R chiralities. To isolate the two chiralities we define for $j=0,2,4,...$ even, the independent canonical fermion operators
\be
\psi_-(j)=\frac{1}{\sqrt{2}}\left(\psi_j+\psi_{j+1}\right)\;,\;
\psi_+(j)=\frac{1}{\sqrt{2}}\left(\psi_j-\psi_{j+1}\right)\,.
\ee
 
In the continuum limit $\psi_+$ will select only the first component of the modes and $\psi_-$ the second one. Hence, using $x= 2 j a$ and $k \rightarrow k a$, $m\rightarrow m a$, taking the $a\rightarrow 0$ limit,  and properly normalizing the modes, obtains
\bea
\psi_+(t,x)= \int \frac{dk}{\sqrt \pi}\, e^{-i k (t-x)}\, d_k \,,\nonumber \\
\psi_-(t,x)=\int \frac{dk}{\sqrt \pi}\, e^{-i k (t+x)}\, R(k) d_k\,,
\eea
with $\{d_k,d_{k^\prime}^\dagger\}=\delta(k-k^\prime)$, and now $k$ extends from $-\infty$ to $\infty$. The vacuum state is $d_k|0\rangle=0$ for $k>0$ and $d_k^\dagger|0\rangle=0$ for $k<0$. 

In the continuum limit, we get for the reflection coefficient
\be
R(k)=\frac{1+i \frac{m}{2k}}{1-i \frac{m}{2k}}=e^{i 2 \delta(m/k)}\,.
\label{420}
\ee
(Note that $R(k)=R^*(-k)$, which reflects charge conjugation symmetry). Furthermore, from (\ref{eq:eta1}), the impurity field is related to the bulk field by
\be\label{eq:eta2}
\eta(E) = - \frac{i}{2 \sqrt \pi}\frac{m^{1/2}}{E}(1+R(E)) d_E\,,
\ee
where $E=k$ is the energy.

This models illustrates very simply the general discussion in \S \ref{subsec:boundaryRG} of left and right movers, and the effect of the boundary in producing a reflection coefficient for the right movers. We have a nontrivial boundary RG flow, and this is reflected in the momentum dependence of $R(k)$. 
For large momentum the phase is $1$ and in the IR is goes to $-1$. 
We conclude that the lattice model realizes a boundary RG flow between
\be
\psi_+(0) = \psi_-(0)
\ee
in the UV, and
\be
\psi_+(0)=-\psi_-(0)
\ee
in the IR.

\subsection{Thermal entropy}\label{subsec:thermal}

We now study the free Kondo model at finite temperature, with the aim of obtaining the thermal boundary entropy.
Let us put the system in a box of length $L$ ($L/a$ sites). We choose the matrix $M$ with $N+1$ sites (including the impurity site $j=-1$, or $N$ without the first site), with $N$ even, $L=N a$, and impose the boundary condition
\be
\psi_{j=N}(k)=0\,, \; \sin(k (N-1) -\delta(k))=0\,. 
\ee
We have written the reflection coefficient $b_k/a_k=e^{i 2 \delta}$. The eigenvalues are then quantized as
\be
k (N-1)-\delta(k)= q \pi\,, \label{defas}
\ee  
giving a spectrum that is  still symmetric with respect to the origin (as implied by charge conjugation symmetry),
which has $N+1$ eigenvalues between $-\pi/2$ and $\pi/2$ for finite $m$. This is not the case in the IR limit $m\rightarrow\infty$, where the quantization condition (\ref{defas}) gives only $N$ eigenvalues. This missing eigenvalue will translate into a running of the impurity thermal entropy in the continuum limit of amplitude $\log2$.

We put the system at inverse temperature $\beta$. The entropy per mode at zero chemical potential is
\be
s(x)=\log(2 \cosh(x/2))-x/2 \tanh(x/2)\,, \hspace{1cm}x=\beta E\,.
\ee 
This is symmetric around $E=0$.  In the limit of large $L$ the modes have small separation $\Delta(k)\sim \pi/N=a \pi/L$ ($\delta$ is a slowly varying function) and the sum can be approximated by an integral
\be
S=\sum_{k} S(k \beta)= \frac{L}{\pi}\int_{-\infty}^\infty dk\, s(k \beta)+ O(1)+O(1/L)=\frac{\pi}{3} TL +O(1)+O(1/L)\,.
\ee
The constant term in the limit $L \to \infty$ defines the thermal boundary entropy.

To get the $O(1)$ term we note that the change of each $k$ due to $\delta$ is a small number $\delta/N=\delta a/L$ and we can put in the infinite $L$ limit
\be
 \Delta S=\frac{\beta}{\pi}\int dk\, s^\prime(\beta k) \delta=\frac{1}{\pi}\int dx\, s^\prime(x) \delta(x,\mu)=\frac{1}{\pi}\int dx\, s(x) G(x,\mu)     
\ee
where $\mu=\beta m$ and 
\be
R(x,\mu)=\frac{2x+i \mu}{2 x-i \mu}\,,\hspace{.6cm}
G(x,\mu)=\frac{-i dR(x,\mu)/dx}{2 R(x,\mu)}=\frac{2 \mu}{\mu^2+4 x^2}\,.
\ee

Note this formula is independent of an overall constant in the reflection coefficient $R$. What matters is the relative dephasing as we move $k$. 

For very small $\mu$, the UV fixed point, or large temperature, we have $\Delta S=\log(2)$, since $G$ goes to a delta function. For large $\mu$, the infrared, $G$ goes to zero and the constant term in the entropy vanishes.    

\begin{figure}[h]
\begin{center}  
\includegraphics[scale=1.2]{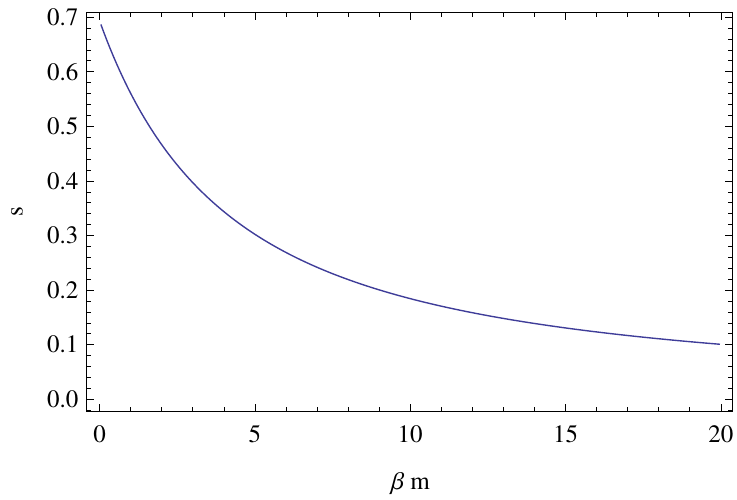}
\captionsetup{width=0.9\textwidth}
\caption{The thermal entropy as a function of $\beta m$.  \label{thermal}}
\end{center}  
\end{figure}

To see more clearly the origin of the monotonicity of the running with temperature (see Figure \ref{thermal}) we compute
\be
\frac{d \Delta S}{d\beta}=\frac{1}{\beta \pi}\int dx\, x s^\prime(x) G(x,\mu)=  -\frac{1}{\beta \pi}\int dx\, \frac{x^2}{2 \cosh^2[x/2]} G(x,\mu)\,.\label{betapri}
\ee
$R$ is the reflection coefficient (\ref{420}) and is independent of the temperature. On the other hand, $G$ depends on $\beta$ leaving the combination $G\ dx$ independent of $\beta$. 
Hence, the boundary entropy decreases monotonically with decreasing temperature (decreases with increasing beta) because $G>0$.

\section{Quantum entanglement in the free Kondo model}\label{sec:application}

In this last section we compute various quantities from quantum information theory in the free Kondo model. Specifically, we focus on the impurity entropy, relative entropy and mutual information. These calculations serve to illustrate in a simple setup the general discussion of \S \ref{sec:relative} and  \S \ref{sec:mutual}.

\subsection{Modular Hamiltonian for spatial intervals}

In \S \ref{subsec:relative-real} we argued that the relative entropy on spatial intervals distinguishes $\rho$ (the full system with the impurity) and $\rho_0$ too much. The contribution from the modular Hamiltonian is expected to grow linearly with the size of the interval, masking the monotonicity of the boundary entropy. We have checked numerically that the expectation value of the modular Hamiltonian indeed grows linearly on the lattice free Kondo model. We now explore another possibility to deal with the contribution of the impurity directly in the continuum, and our conclusion will be again that on the spatial interval relative entropy is too large. 

In order to decouple the localized impurity term in $\langle T_{00}\rangle$ and better understand the different contributions as we approach the boundary, it is useful to regularize around the boundary, and consider an interval $x_1 \in (\delta, r)$, with $\delta \to 0$. It is also convenient to work with the equivalent theory of a single chiral fermion $\psi_+$ along the full line, by reflecting $\psi_+(x_1) = \psi_-(-x_1)$ for $x_1<0$.\footnote{This gives a continuous wavefunction at short distances, since the UV boundary condition imposes $\psi_+(0)=\psi_-(0)$.} Therefore, we need to calculate the modular Hamiltonian for a free fermion on a region formed by two intervals,
\be
A=(-r,-\delta)\cup (\delta,r)\,.
\ee
Now the impurity falls outside of the region and we can assume the stress tensor vanishes inside $A$. We have to determine the behavior of $\Delta\langle \mathcal H\rangle$ as a function of $r$ when $\delta \to 0$.

For two intervals the modular Hamiltonian contains non-local terms, and operators other than the stress tensor. 
The modular Hamiltonian for the two interval region $A$
 for a massless $d=2$ Dirac field is \cite{Casini:2009vk}
\be
\mc H= \mc H_{loc}+\mc H_{nonloc}\,.
\ee
The local part is (we write $x \equiv x_1$ below)
\bea
 \mc H_{loc}&=&2 \pi\int_A dx\, f(x) T_{00}(x)\,,\label{local11}\\
f(x)&=&\frac{(x^2-\delta^2)(r^2-x^2)}{2(r-\delta)(x^2+\delta r)} \,.
\eea
For a fixed point $x$, the local term converges to the one of an interval of size $2r$ when $\delta\ll r$, that is $f(x)\rightarrow (r^2-x^2)/(2r)$.\footnote{Note that $f(x)$ vanishes linearly at the boundaries, where the modular Hamiltonian is ``Rindler like''. } However, this regularized expression is insensitive to the impurity stress tensor, which is localized at $x=0$.

\begin{figure}[h]
\begin{center}  
\includegraphics[scale=0.8]{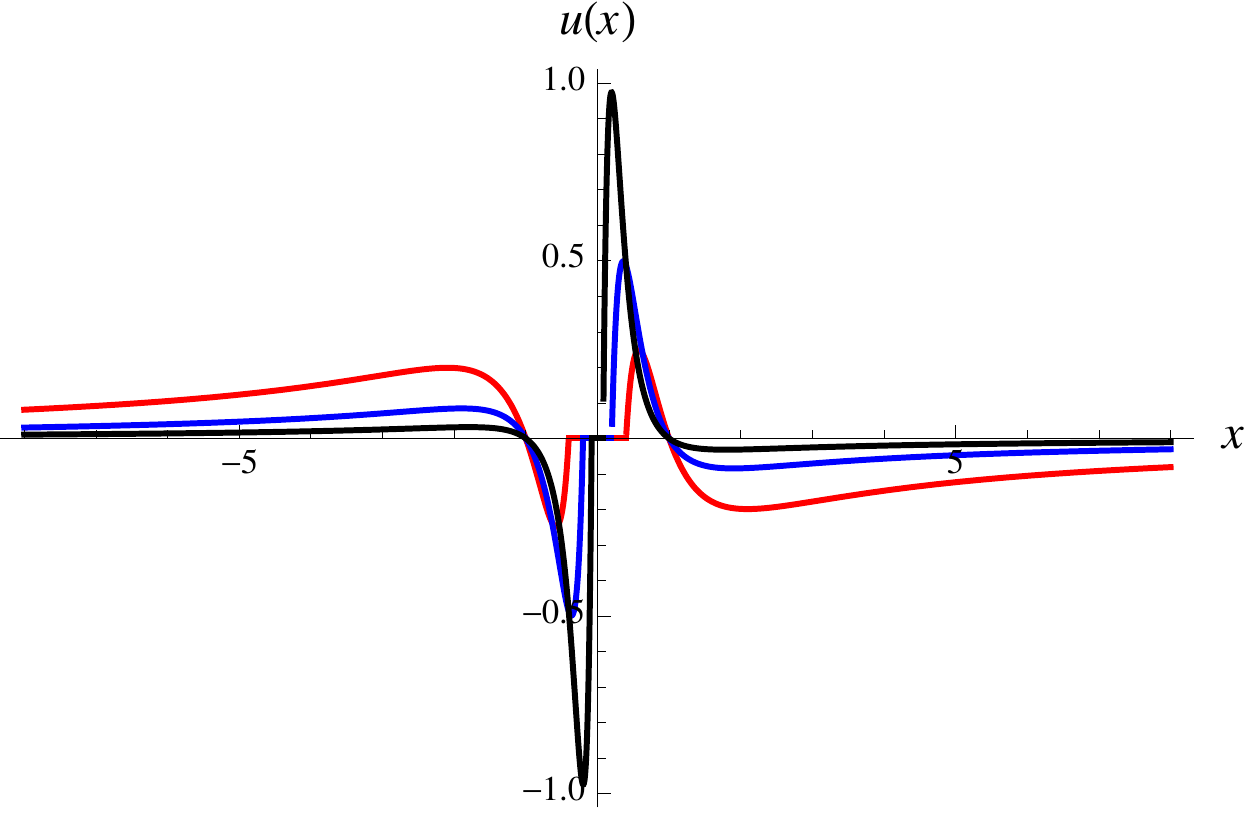}
\captionsetup{width=0.9\textwidth}
\caption{The function $u(x)$ multiplying the non local term for $r=1$, $\delta=0.4,0.2,0.08$, in red, blue and black, respectively. \label{fig:nonlocal}}
\end{center}  
\end{figure}

To write the non-local part define the following global conformal transformation
\be
\bar{x}=-\frac{r\delta }{x}\label{reflec}\,.
\ee
This maps one interval into the other. We have
\bea\label{eq:nonloc}
\mc H_{nonloc}&=&-\pi i \int_A dx\,  u(x) \psi_+^\dagger(x)\psi_+(\bar{x})\\  \nonumber
u(x)&=& \frac{r\delta }{ (r-\delta)}\frac{(r^2-x^2)(x^2-\delta^2)}{x(x^2+r\delta )^2}\,.
\eea
Only $\psi_+(x)$ appears here, as opposed to a full Dirac fermion, because the theory on the full line contains only a chiral fermion.
Note $u(x)$ goes to zero with $\delta$ for any fixed $x$ but develops larger peaks near the origin as $\delta \to 0$ (see Figure \ref{fig:nonlocal}). This structure will be responsible for the linear in $r$ dependence of $\Delta \langle \mc H \rangle$.

The local part of the modular Hamiltonian does not contribute to $\Delta\langle \mc H\rangle$ because the expectation value of $T_{00}$ is zero in $A$. Then we examine the non local term in the limit of $\delta \rightarrow 0$. 
The expectation value of $\mc H_{nonloc}$ involves the fermion correlator with points on opposite sides of the impurity, i.e., the two-point function between the left and right movers in the model defined on $x>0$. These differ by the reflection factor $R_m(k)$. The system without the impurity is obtained for $m=0$, so
\be
\Delta\langle \mc H \rangle=i\int_\delta^r dx\,u(x)\,\int_0^\infty dk\,(R_m(k)-R_{m=0}(k))e^{-i k(x+r\delta/x)}\,.
\ee
Here we used the plane wave solutions of \S \ref{sec:freeKondo}, and the fact that the equal time fermion correlator in momentum space projects on the positive energy states $k<0$. The integral over $x$ is dominated by the behavior of $u(x)$ near the maximum $x \sim (\delta/r)^{1/2} r$, and hence it is sufficient to approximate the reflection factors by their UV behavior (equivalently, by an expansion around $m=0$). As shown in Appendix \ref{app:modular}, this leads to
\be
\Delta\langle \mc H \rangle \sim \,m\,r\,\log(m (r\delta)^{1/2})\,.
\ee

This shows the linear dependence in $r$ for the expectation value of the modular Hamiltonian. It also exhibits a logarithmic divergence (consistent with (\ref{consist})) as the cutoff $\delta$ is removed --the two-interval result does not converge to the single-interval answer. It is associated to UV modes localized near the impurity, which contribute to the entanglement.

\subsection{Kondo model on the null line}
\label{subsec:null}

\begin{figure}[h!]
\begin{center}  
\includegraphics[scale=0.55]{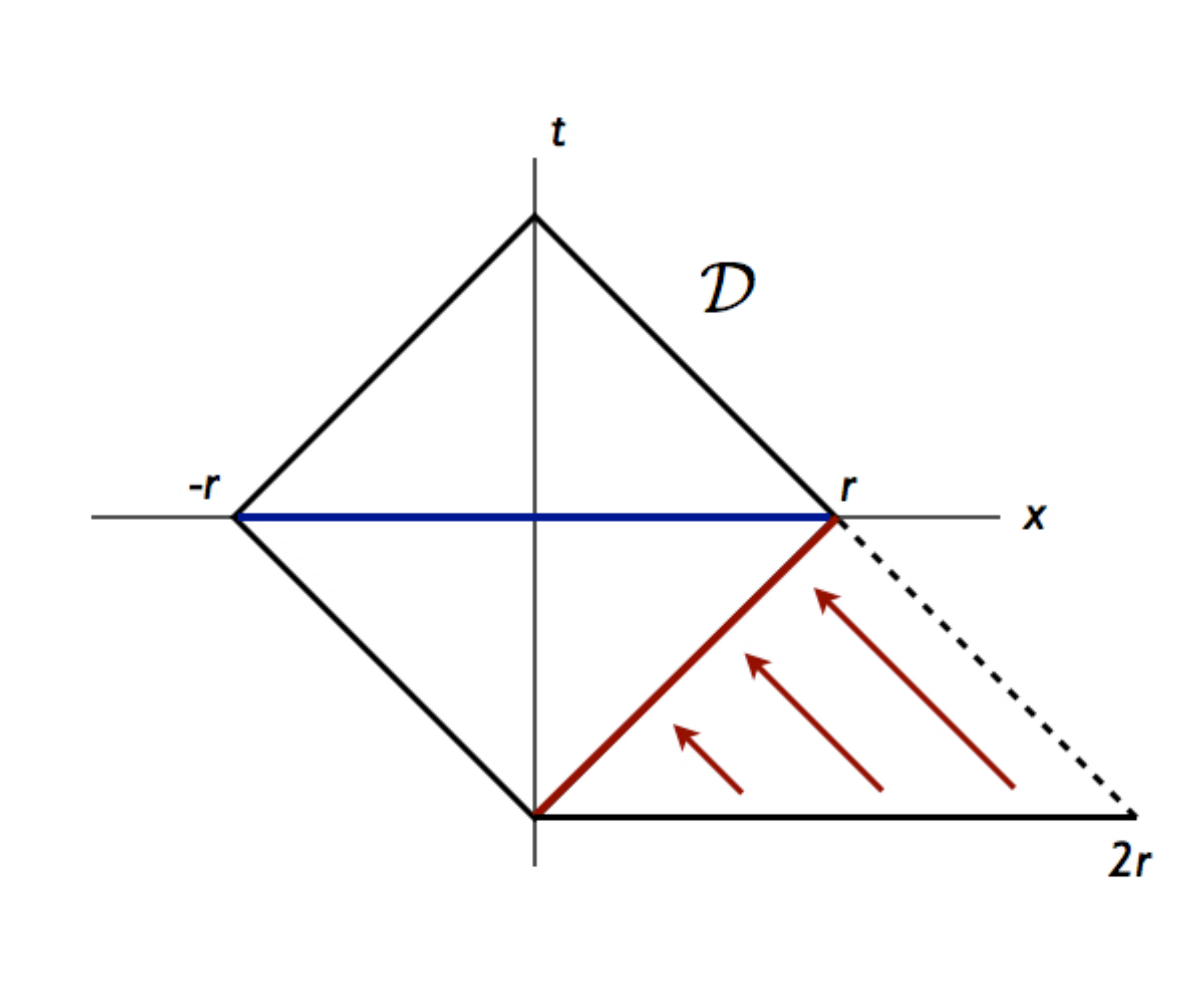}
\captionsetup{width=0.9\textwidth}
\caption{In the model the two chiralities on the half plane can be unfolded into a single chirality on the whole plane.  Since the evolution is unitary, the entanglement entropy for the blue segment is the same as for the red segment, as well as on the black segment of length $2r$ that touches the impurity at its left endpoint.}
\label{fig:causal3}
\end{center}  
\end{figure}

Let us now consider the Kondo model on null segments; the setup is shown in Figure \ref{fig:causal3}. In this chiral model, the nontrivial two-point functions on null segments are the same as in the theory without impurity because only one chirality contributes. As a result, $\Delta \langle \mc H \rangle =0$.

The lattice calculations cannot be done at time $t=0$ by keeping only the combination $\psi_j+\psi_{j+1}$ in the algebra, which in the continuum limit is proportional to $\psi_+(x)$. This is because a large, volume increasing entropy will be generated by non vanishing entanglement with the other components $\psi_j-\psi_{j+1}$. This entanglement is not present in the continuum fields. In other words, the spatial lattice is a bad regularization to make calculations on the null line. 

However, we can do calculations directly on the null line in the continuum limit in the present case. The correlators for a null interval have the form of a kernel 
\be
C=\left(
\begin{matrix}
1/2 &  \langle \eta \psi_+^\dagger(x)\rangle\\
\langle \eta \psi_+^\dagger(y)\rangle^* &   C_0(x,y)
\end{matrix}
\right)
\ee
where we have used $\langle \eta \eta^\dagger\rangle=1/2$, and $C_0(x,y)=\langle \psi_+(x) \psi_+^\dagger(y)\rangle$ coincides with the corresponding correlator in the free Dirac model without impurity. The impurity establishes correlators with the field without changing the correlator that the chiral field has with itself in the bulk without impurity. It is interesting to note this would not have been possible if the state of the field had been pure, since a pure state cannot have correlations with exterior systems. It is the reduction to the half line that allows correlations with the impurity. The effect of the impurity will be just to slightly purify (by a $\log(2)$ amount) the field state in the half line.

The correlator of $\eta$ and $\psi_+$ follows from (\ref{eq:eta2}), and reads
\be\label{eq:etapsi}
\langle \eta \psi_+^\dagger(x)\rangle=\int_0^\infty \frac{dE}{2 \pi}\,\left( -\frac{i}{2} \frac{m^{1/2}}{E}\right)(1+R(E)) e^{-i E x}=\frac{i m^{1/2}}{2 \pi} e^{m x /2} Ei(-m x/2)\,,
\ee  
where $Ei(x)$ is the exponential integral function $Ei(x)=-\int_{-x}^\infty dt\,e^{-t}/t$.

The kernel $C_0(x,y)$ in an interval $(0,r)$ can be diagonalized \cite{Casini:2009vk}, with eigenfunctions and eigenvalues
\be
\int_0^r dy\, C_0(x,y) \psi_s(y)=  \lambda_s \psi_s(x)\,,
\ee
where
\bea
\lambda_s&=&\frac{1}{2}(1+\tanh(\pi s))\,,\\
\psi_s(x)&=&\frac{r^{1/2}}{(2\pi)^{1/2}(x(r-x))^{1/2}}e^{-i \log(x/(r-x))}\,. \nonumber
\eea
The eigenfunctions are normalized with respect to the parameter $s\in (-\infty,\infty)$ 
\be
\int_0^r dx\, \psi_s(x) \psi_{s^\prime}^*(x)=\delta(s-s^\prime)\,.
\ee
Using this basis we can rewrite the correlator as 
\be
C=\left(
\begin{matrix}
1/2 &  a(s,mr)\\
 a^*(s,mr)&   \textrm{diag}(\lambda_s)
\end{matrix}
\right)\label{ccc}
\ee
where 
\be
a(s,mr)=\int_0^r dx\, \psi_s(x) \langle \eta \psi_+^\dagger(x)\rangle=\int_0^{mr} dz\, \frac{i (mr)^{1/2}}{(2\pi)^{3/2}} \frac{e^{z/2}Ei(-z/2)}{z^{1/2}(mr-z)^{1/2}}e^{-i s \log\frac{z}{mr-z}} \,.
\ee

Now we compute the relative entropy of the state determined by this correlator and the UV fixed point $m=0$. This is a decoupled state $\rho^0=\rho^0_F\otimes\rho_{imp}^0$, where $\rho^0_F$ is the density matrix of the fermion system without impurity. Note that the state $\rho$ reduced to algebra of the impurity, or to the field, exactly coincides with $\rho^0$. Hence $\Delta \mc H=0$. We have 
\be
S_{\textrm{rel}}=-\Delta S=S(\rho^0_F)+S(\rho_{imp})-S(\rho)=I(imp,F)\,,
\ee
which coincides with the mutual information between the impurity and bulk system on the null line. 

To evaluate this we use the expression 
\be\label{eq:Sfermion}
S(r)= -\,\tr\left(C\,\log C+(1-C) \log(1-C) \right)
\ee 
for the entropy of a free fermion in terms of the correlator (\ref{ccc}). It is convenient to write this entropy as an integral expression in terms of determinants \cite{Casini:2009vk},
\be
S=2\int_1^\infty d\lambda\, \frac{\log \det(\lambda^{-1/2}(1+(\lambda-1) C))}{(\lambda-1)^2}\,. \label{dets}
\ee 
In the basis of expression (\ref{ccc}) this determinant has a simple form as can be seen expanding it by the first column. For a matrix of the form 
\be
N=\left(
\begin{matrix}
a &  \vec{b}\\
 \vec{b}^* &   \textrm{diag}(c_i)
\end{matrix}\right)
\ee
we have
\be
\det(N)=\det(c)\left(a-\sum_i \frac{|b_i|^2}{c_i}\right)\,.
\ee
\begin{figure}[!h]
\begin{center}  
\includegraphics[scale=0.6]{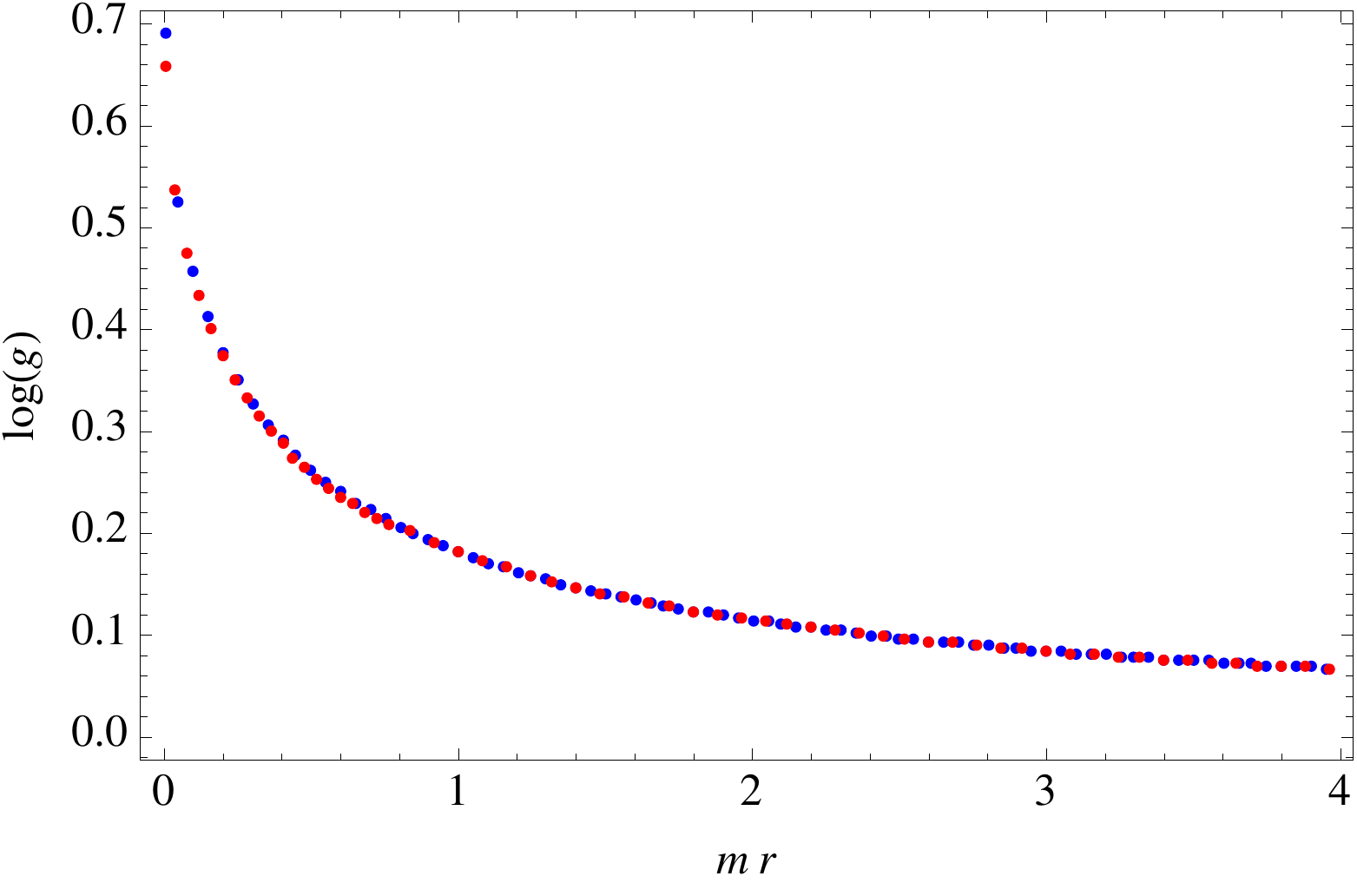}
\captionsetup{width=0.9\textwidth}
\caption{$\log(g)$ as a function of $mr$ computed analytically in the null interval (Blue) using (\ref{nullint}) and on the spatial interval (Red) using a lattice regularization.   \label{fig:logg}}
\end{center}  
\end{figure}

Using this we see the entropy of the unperturbed field decouples in (\ref{dets}) and we get the exact analytic expression for the function $\log g(r)=S(\rho(2r))-S(\rho^0_F(2r))$. We have to use entropies for the interval in null coordinate of a size $2R$ since this is the one that is unitarily mapped to a spatial interval os size $r$ in the half place, see figure \ref{fig:causal3}. We have 
\bea
\label{nullint}
\log g(r)&=&2\int_1^\infty d\lambda\,\frac{1}{(\lambda-1)^2}\\
& \times&\log\left(\frac{1}{2}(\lambda^{1/2}+\lambda^{-1/2})-\frac{(\lambda-1)^{2}}{\lambda^{1/2}}\int_{-\infty}^\infty ds\, \frac{|a(s,2mr)|^2}{1+\frac{1}{2}(\lambda-1)(1+\tanh(\pi s))}\right)\nonumber \,.
\eea  
The integrals can be exactly evaluated for $mr\rightarrow 0$ and $mr\rightarrow \infty$ giving $\log(2)$ and $0$ respectively, as expected. For intermediate values, a numerical evaluation of the integral shows that the result on the null line coincides with the continuum limit of $g(r)$ evaluated on the lattice on spatial intervals (see figure \ref{fig:logg}). 

Figure \ref{fig:logg} shows the numerical evaluation of $g(r)$. We compare the results on the null line coming from the numerical integration of  (\ref{nullint}) and the results on a spatial interval. We calculate entanglement entropies numerically for the lattice model of \S \ref{subsection:lattice}, take the continuum limit and use (\ref{eq:Sfermion}). The boundary entropy $\log g$ is defined on the lattice as the difference of entanglement entropy for the system with impurity (the $\eta$ field of \S \ref{subsection:lattice}) and without impurity; see (\ref{eq:gdef1}) for an alternative definition.

We have done the numerical calculations using two different methods and checked their agreement. First, we use a finite lattice, diagonalize the Hamiltonian and then compute the fermion correlators $C_{ij}$ restricted to the interval of interest. In this manuscript we used lattices from $2000$ sites up to $8000$ sites. The other method is to work on an infinite lattice, and use the wavefunctions obtained in \S \ref{subsection:lattice} to calculate the equal time correlators
\be
C_{ij}=- \int_0^{\pi/2}\,\frac{dk}{2\pi} \psi_i^\dag(k) \psi_j(k)\,.
\ee
This uses the fact that the correlator in momentum space at $t=0$ is a projector on positive-energy states.
More details of this procedure for free fields may be found in e.g. \cite{Casini:2009sr}.

The relative entropy given by (\ref{eq:gteo}) reads 
\be
S_{\textrm{rel}}=\log(g(0)/g(r))\,,
\ee
and is indeed positive and increasing. Here we have checked that $\Delta \langle \mc H \rangle=0$ for our free model.


\subsection{Mutual information in the free Kondo model}\label{subsec:mutualfree}

Finally, we evaluate the mutual information in the free Kondo model. This will characterize the correlations between the impurity and bulk systems studied in \S \ref{subsec:Icorrelations} as well as the function $\t g(r)$ obtained in \S \ref{subsec:general-proposal}.

\begin{figure}[!h]
\begin{center}  
\includegraphics[scale=0.47]{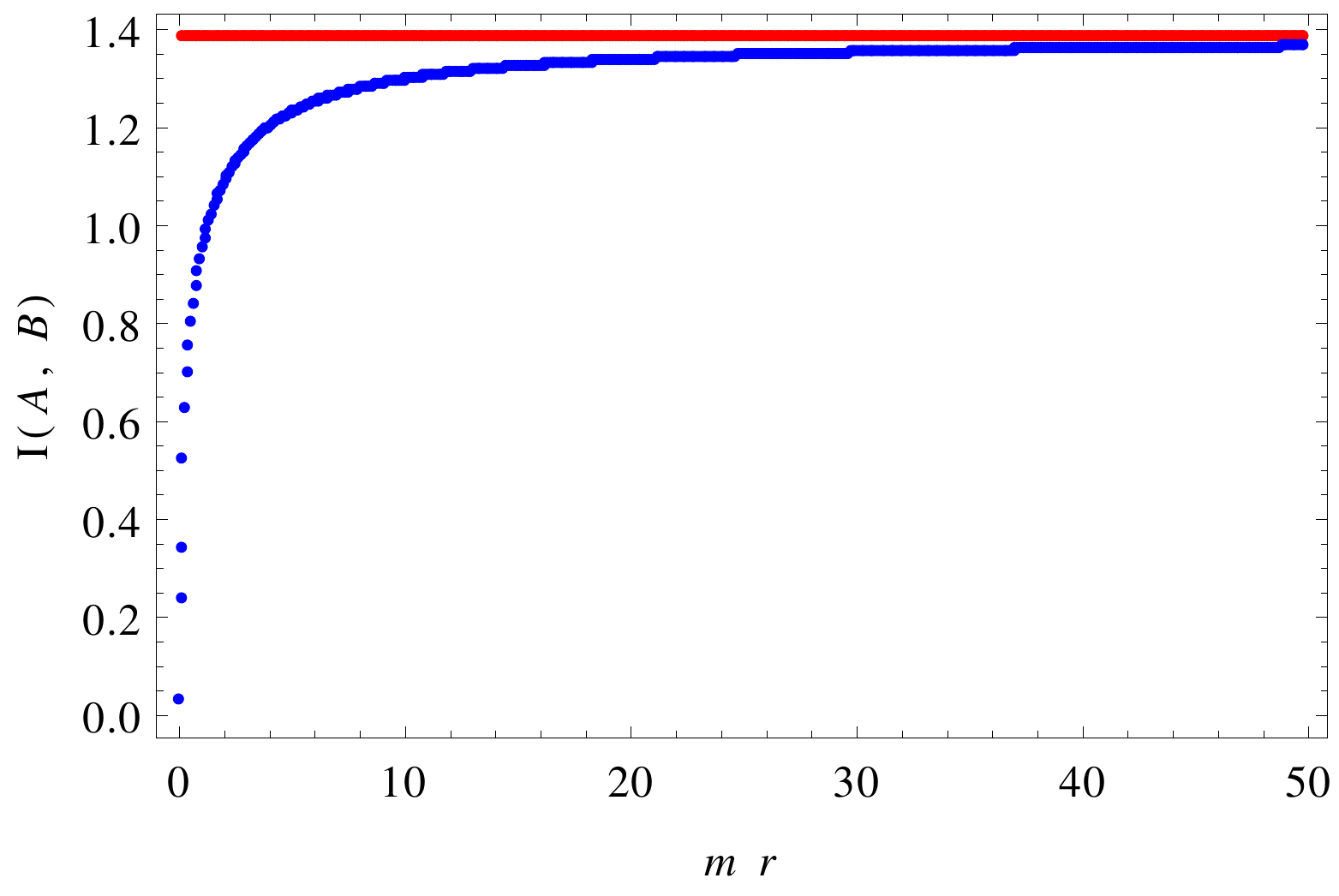}\hfill\includegraphics[scale=0.47]{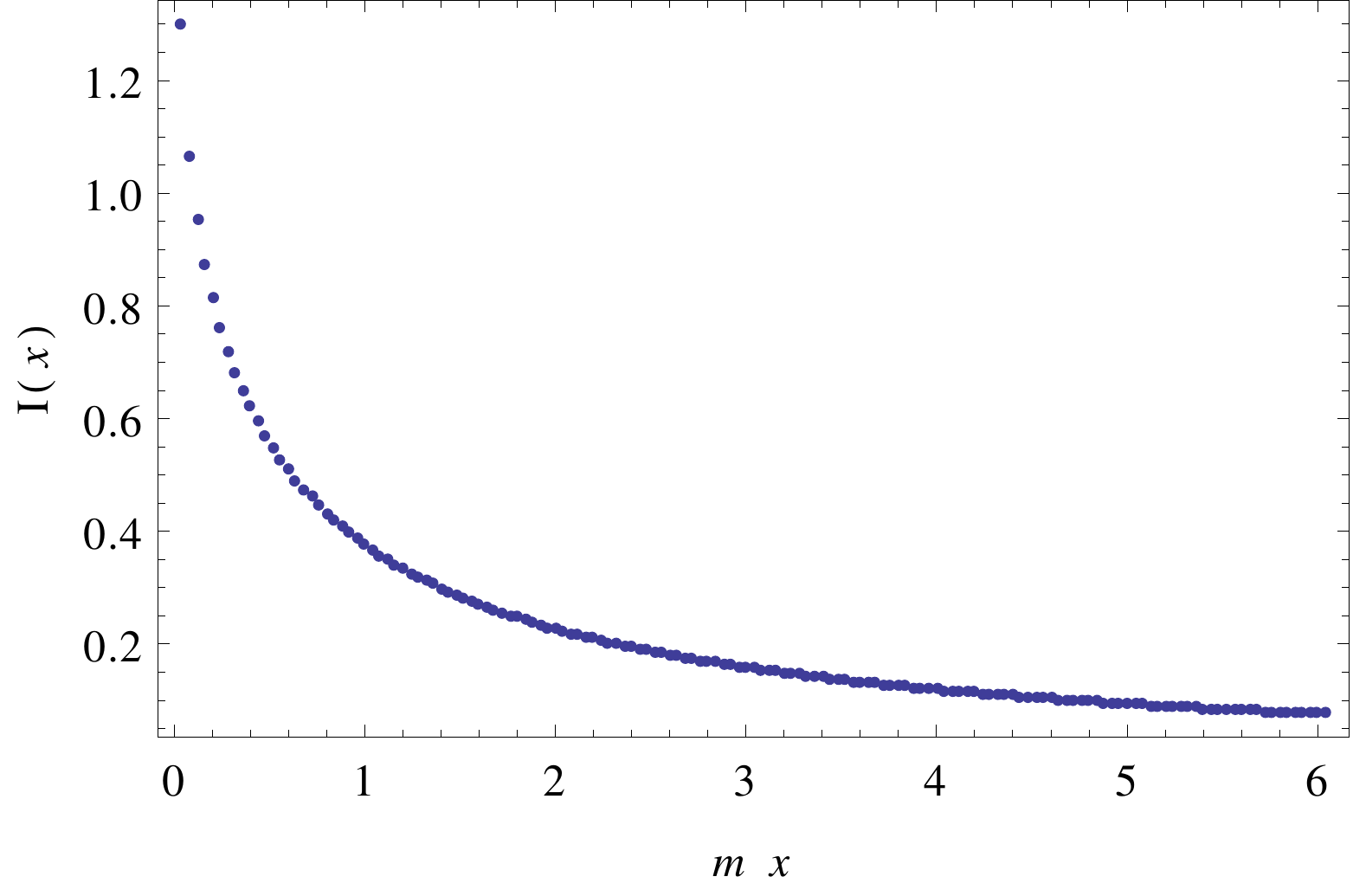}
\captionsetup{width=0.9\textwidth}
\caption{Left: Mutual information between the impurity and bulk fermions as a function of $m r$. For large $r$, $I(r) \to 2 \log 2=1.386...$, twice the value of the total impurity entropy (shown in red). Right: Mutual information between the impurity and a far-away interval of fixed size. In particular for this plot we chose as region $B$ the interval $[m x ,m x +10].$\label{fig:mutual1}}
\end{center}  
\end{figure}

The numerical result for the mutual information $I(A,B)$ is presented in the left panel of Figure \ref{fig:mutual1}, with $A$ the interval $[0,\epsilon]$, containing the impurity and $B=[\epsilon^\prime,r]$. The mutual information starts at zero, and then grows with the size of the interval; it asymptotes to $2 \log 2$ for large $r$. It is very well approximated by
\be
I(r) \approx 1.388-\frac{0.98}{r}+\frac{1.01}{r^2}\,.
\ee
The asymptotic value is $\sim 2 \log 2$, twice the value of the total impurity entropy, in agreement with the previous discussion. The subleading powers of $r$ encode information about correlators in the theory, as we discuss shortly. We see that our proposal for the mutual information indeed captures the entanglement between boundary and bulk, and is free of UV divergences. The mutual information in this case reflects the boundary RG flow between the $+$ boundary condition for $m r \ll 1$, and the $-$ boundary condition for $m r \gg 1$.

Next, we may also use the mutual information to characterize the boundary-bulk correlations, as argued in \S \ref{subsec:Icorrelations}. For this, let $A$ be an infinitesimal interval containing the impurity, and choose $B$ as an interval of fixed size, at a distance $x$ from the origin. In the limit of large $x$, the twist operators that implement the cuts in the EE may be approximated by local QFT operators, and hence we expect 
$I(x) \approx \frac{1}{|x|^{2\Delta}}$
with $\Delta$ the smallest operator dimension that contributes to the OPE of the twist operators \cite{Cardy}. We plot this quantity in the right panel of Figure \ref{fig:mutual1}. 

This result is very well fitted by $c_0+c_1/x+c_2/x^2$, and hence the leading correlator that contributes to the mutual information has $2\Delta=1$. This is the behavior we expect from the correlation (\ref{eq:etapsi}) between the impurity and bulk fermions, 
\be
\langle \eta(t) \psi_+(x,t) \rangle = \int_0^\infty\,\frac{dE}{2\pi}\,\eta(E) \psi_+(E,x) \to\frac{1}{m^{1/2}|x|}\,.
\ee
This also reflects the fact that the physical dimension of $\eta$ in the IR is $\Delta=1/2$, while in the UV we had $\Delta=0$; see (\ref{eq:Seffeta}). The mutual information nicely captures this flow.

\begin{figure}[!h]
\begin{center}  
\includegraphics[scale=0.47]{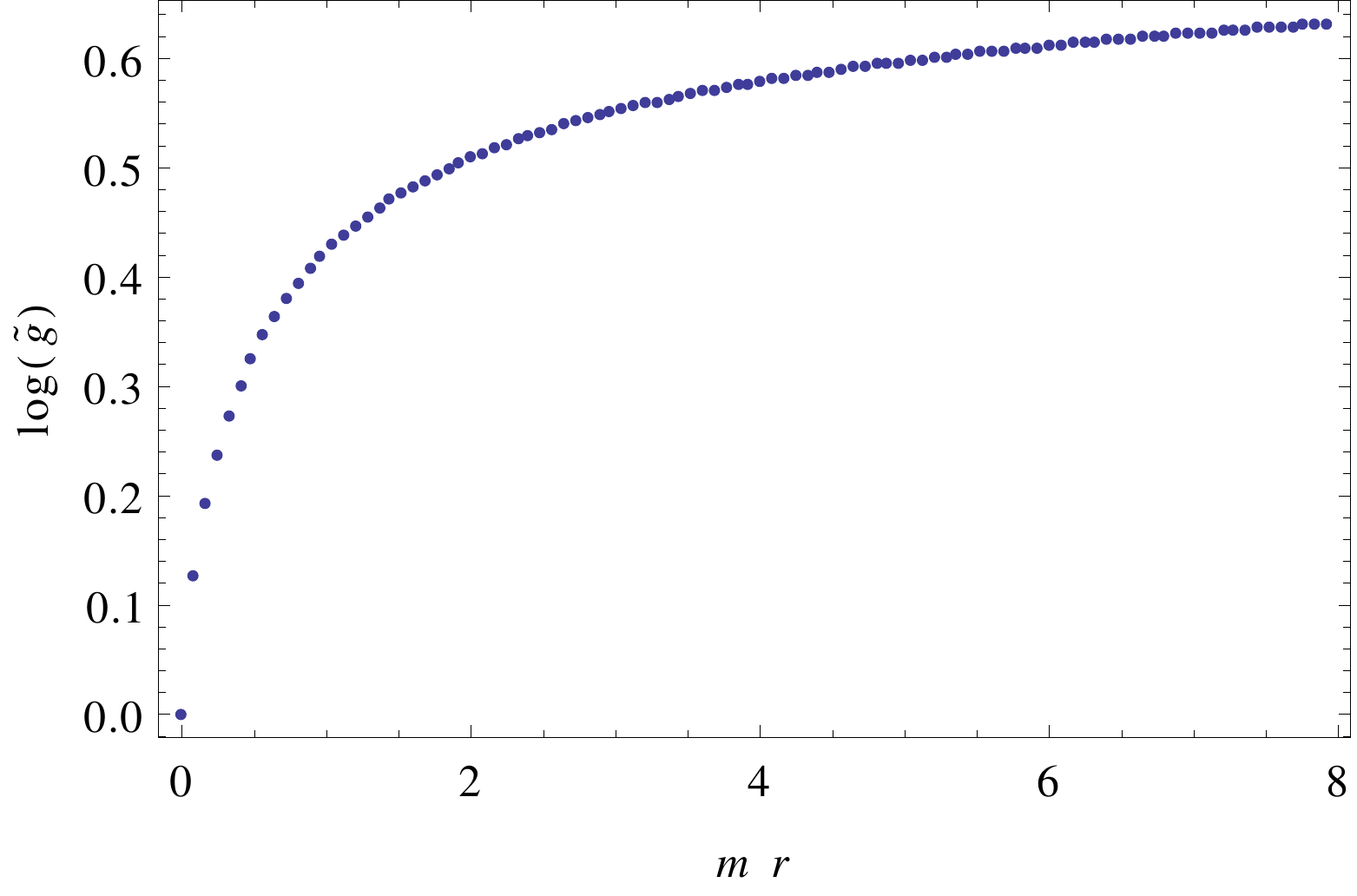}\hfill
\includegraphics[scale=0.47]{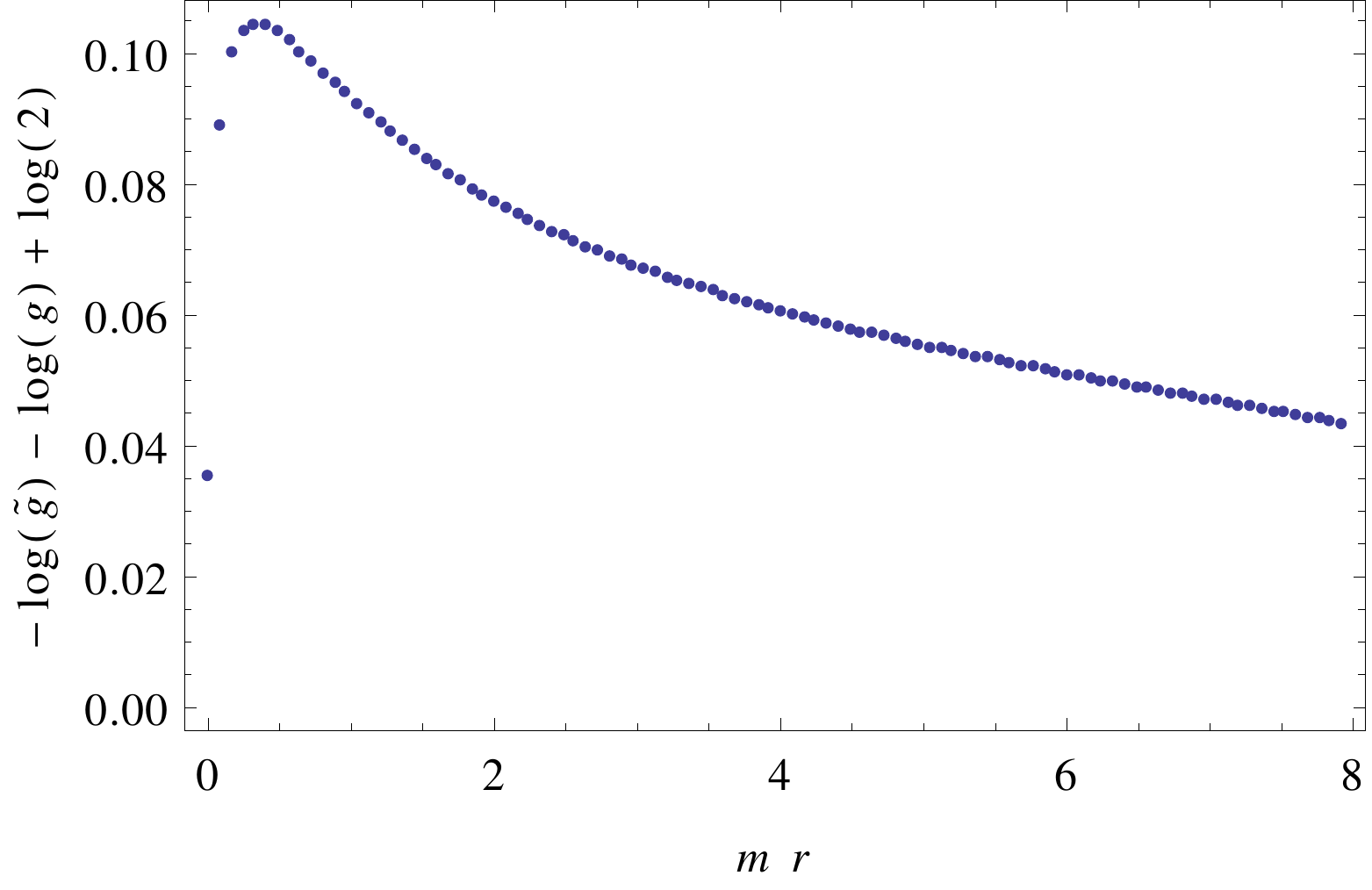}
\captionsetup{width=0.9\textwidth}
\caption{Left: $\log \tilde{g}$ as a function of $mr$ \label{tildeg}. Right: $\log2-\log g-\log \tilde g$ as a function of $mr$.}
\end{center}  
\end{figure}

Finally, we compare the functions $g(r)$ and $\t g(r)$, defined as discussed in \S \ref{subsec:general-proposal},
\bea
- \log\,g(r)&=&S(A \cup B)-S_\text{no imp}(B) \nonumber\\
 -\log\,\tilde g(r)&=&S(B)  - S_\text{no imp}(B)\,.
\eea
The left hand side of Figure \ref{tildeg} shows how $\log \tilde g$ approaches $0$ for small values of $r$ and asymptotes to $\log 2$ as $r\rightarrow\infty$.  The behavior of $\log g$ was given before in Figure \ref{fig:logg}. We also checked that the combination $\log2 +\log\,\tilde g(r)- \log\,g(r)$  - given in  (\ref{Irlogg}) - coincides within numerical precision with $I(A,B)$, a magnitude with a well-defined continuum limit. 

It is also interesting to consider the magnitude $\log2-\log g-\log \tilde g$, as a way of characterizing the Kondo cloud. This is zero for the probabilistic model presented in \S \ref{subsec:valence}. As we can see from the right hand side of  Figure \ref{tildeg}, in the free model the quantity $\log2-\log g-\log \tilde g$ approaches zero for $r\rightarrow 0,\, \infty$, but has a nontrivial profile for finite values of $r$. The plot has a maximum around $m r\approx0.5$, which agrees with the expectation that the Kondo cloud should be of order $mr\approx1$. As a possible future direction, it would be interesting to find a well defined continuum information theory quantity that characterizes the Kondo cloud.

\section{Conclusions and future directions}\label{sec:final}

In this work we proved that the boundary entropy of BCFTs is a relative entropy, and that it decreases under boundary RG flows. This is the first monotonicity theorem where we have a quantum information understanding of RG flow. Here we see the decrease on $g(r)$ as a result of increased distinguishability between the state with the boundary perturbation and the UV CFT vacuum towards the IR, as we allow more low energy operators to be used to discriminate between these two states. The effect of the boundary RG flow is to change correlations and increase distinguishability. We can also rephrase this saying that the state $\rho$ of the CFT with boundary RG flow looses information about the UV fixed point as we go towards the IR, that is, it is less able to faithfully reproduce the UV CFT vacuum.   

More generally, we argued that methods from quantum information theory provide a valuable understanding of boundary RG flows. Besides the relative entropy, we focused on the mutual information between the impurity and bulk degrees of freedom, and how it encodes correlations. We illustrated these results in terms of a new solvable Kondo model of relativistic free fermions.

We would finally want to discuss some future directions suggested by these results. We have seen that working in the null basis provides important simplifications, and it will be interesting to see how this works out for other quantum information measures in the presence of impurities. Our methods may also have applications to C-theorems. It would also be important to try to generalize our approach to higher dimensions and different defects. Recent work in higher dimensions includes \cite{Gaiotto:2014gha, Jensen:2015swa}. Moreover, it would be interesting to explore the connection with holographic results on defect entropy \cite{Azeyanagi:2007qj, Fujita:2011fp, Nozaki:2012qd, Estes:2014hka, Erdmenger:2015spo}. Finally, if a physical realization of our free Kondo model is possible,\footnote{We thank E. Fradkin for pointing out to us that the free fermion model appears as a special limit of the interacting Kondo problem \cite{free1, free2}.} this would provide a system where measures of quantum information quantities may be easy to perform, in particular, the role of multipartite entanglement and the Kondo cloud could be further clarified.

\section*{Acknowledgments}
We thank J. Cardy, J. Erdmenger, E. Fradkin, R. Myers, E. Tonni for clarifications, discussions and comments. We thank specially E. Witten for suggestions that led to important improvements on the earlier version of this work. This work was supported by CONICET PIP grant 11220110100533, Universidad Nacional de Cuyo, CNEA, and the Simons Foundation ``It from Qubit'' grant.

\appendix

\section{Nonrelativistic Kondo model}\label{app:nonrel}

There is a nonrelativistic model which is closer in spirit to the original Kondo problem, and which reduces at low energies to the previous relativistic setup. Consider a nonrelativistic fermion in $d$ spatial dimensions at finite density, interacting with a fermionic impurity at $x=0$:
\be\label{eq:nonrel}
L= \int\,d^dx \;\psi^\dag \left(i \partial_t +\frac{\nabla^2}{2m}-\mu_F \right) \psi+ \delta^d(x) \left(\eta^* i \partial_t \eta+ m^{1/2}(\eta \psi+c.c.)\right)\,.
\ee
As in the original Kondo model, the spherical symmetry implies that this can be reduced to a one-dimensional problem along the radial direction $r>0$, with the impurity located at $r=0$:
\be
L = \int_0^\infty dr\;\psi^\dag \left(i \partial_t +\frac{1}{2m}\frac{d^2}{dr^2}-\mu_F \right) \psi+\delta(r) \left(\eta^* i \partial_t \eta+ m^{1/2}(\eta \psi+c.c.)\right)\,.
\ee

Due to the Fermi surface at $k_F^2/(2m) =\mu_F$, the nonrelativistic fermion describes two chiralities,
\be
\psi_\pm(x) = \int_{-\Lambda}^\Lambda dk\,e^{\pm i k r}\psi(k+k_F)\,,
\ee
where $\Lambda$ is a momentum cutoff. At half-filling, $k_F=\frac{\pi}{2}\Lambda$, and this model reduces to the previous relativistic theory. This can be seen by discretizing (\ref{eq:nonrel}) on a lattice with $a=1/\Lambda$, which yields a dispersion relation $\varepsilon(k) =-\cos(ka)$. All negative energy states here are occupied, and form the Fermi surface with Fermi momentum $|k_F|= \pi/(2a)$.

\section{Fermion boundary conditions}\label{app:conditions}

This Appendix discusses in more detail the consistent boundary conditions of the fermion theory with boundary. This analysis is well-known, but we have found it useful to include it here for completeness.

Since the bulk lives in the half space, a consistent boundary condition comes from imposing the vanishing of the boundary term in the action variation:
\be
\delta S_{bdry}=\int dx_0\,i(\psi_+ \delta \psi_+^*-\psi_-^* \delta \psi_-)\Big|_{x_1=0}\;\Rightarrow\;\psi_+ \delta \psi_+^*=\psi_-^* \delta \psi_-\Big|_{x_1=0}\,.
\ee
Therefore, the set of consistent boundary conditions are
\be\label{eq:alphacond}
\psi_+(x_0,0)= e^{2\pi i \nu}\psi_-(x_0,0)\,.
\ee
Hence the bulk degree of freedom is a single chiral fermion: the right-mover is determined in terms of the left mover due to the reflection condition on the wall at $x_1=0$. 

The Dirac Hamiltonian $H= i \gamma^0 \gamma^1 \partial_1$ is hermitean for any real $\nu$. To see this, evaluate the hermiticity condition in the half-line for wavefunctions $\psi_1, \psi_2$
\be
\int_0^\infty dx_1 \psi_2^\dag(x) i \gamma^0 \gamma^1 \partial_1 \psi_1(x)=\int_0^\infty dx_1 (i \gamma^0 \gamma^1 \partial_1\psi_2(x))^\dag\, \psi_1(x)\,.
\ee
This requires the vanishing of the boundary term,
\be
\psi_2^\dag \gamma^5 \psi_1 \Big|_{x=0}=0\,,
\ee
where we used $\gamma^0 \gamma^1 = \gamma^5 = \sigma_z$. Writing the fermion wave-functions as $\psi= (\psi_+,\psi_-)$, obtains
\be
\psi_+^\dag \psi_+ = \psi_-^\dag \psi_- \;\Rightarrow\; \psi_+(0) = e^{2\pi i \nu} \psi_-(0)\,.
\ee
Hence this set of boundary conditions preserves the Hermiticity of the operator in the half-line. It can further be checked that these boundary conditions impose $T_{01}=0$ on the line $x_1=0$, and are conformal invariant boundary conditions. 

For a Majorana fermion in 1+1 dimensions, only $\nu=0$ and $\nu=1/2$ are allowed. These are the $+$ and $-$ boundary conditions used in this work, also known as the Ramond and Neveu-Schwarz boundary conditions in the context of string theory. For a Dirac fermion, any real $\nu$ is also allowed. Only $\nu=0\,,\,1/2$ preserve charge conjugation symmetry, which exchanges $\psi_+ \to \pm \psi_-$.\footnote{Recall that in two dimensions charge-conjugation acts on a Dirac fermion as $\psi_C(x)=\gamma^1 \psi^*(x)$ or $\psi_C(x)=\gamma^5 \gamma^1 \psi^*(x)$. In components, this gives $\psi_+ \to \pm \psi_-$, i.e. the two independent fermion creation operators are exchanged. The impurity fermion $\chi$ satisfies the Majorana condition $\chi^*=C \chi$. } 

Although in this work we have restricted to $\nu=0\,,\,1/2$, we note that other values of $\nu$ can be achieved in the lattice model by turning on a chemical potential at the impurity. For instance, adding terms to the Lagrangian of the form $\mu \bar \chi \gamma^0 \chi$, or a delta-function coupling between the bulk and impurity fermions, $\delta(x_1) \mu \bar \chi \gamma^0 \psi(x)$ leads to boundary RG flows with nontrivial $\nu$.  We hope to study these more general RG flows in the future.

\section{Calculation of the modular Hamiltonian}\label{app:modular}

This Appendix presents the calculation of the modular Hamiltonian for two intervals, 
\be
\Delta\langle \mc H \rangle=i\int_\delta^r dx\,u(x)\,\int_0^\infty dk\,(R_m(k)-R_{m=0}(k))e^{-i k(x+r\delta/x)}\,.
\ee
We want to evaluate this quantity in the limit $\delta/r \to 0$.

Let us first simplify the integral over $x$. Working with dimensionless variables $\t x= x/r, \t \delta=\delta/r, \t k = kr, \t m = m r$, we have
\be
\int_\delta^r dx\,u(x)\,e^{-i k(x+r\delta/x)}= r \int_{\t \delta}^1\,d \t x \,u(\t x)\,e^{-i \t k(\t x+\t\delta/ \t x)}\,.
\ee
The function $u(\t x)$ has a maximum at $\t x \sim \t \delta^{1/2}$. In particular, in the limit $\t \delta \to 0$, the maximum is located at $\t x_0= (\t \delta/3)^{1/2}$. This suggests changing variables to $\t x = \t \delta^{1/2} y$. Taking now the limit $\t \delta \to 0$ obtains
\be
\int_\delta^r dx\,u(x)\,e^{-i k(x+r\delta/x)} \approx r \int_{\t \delta^{1/2}}^{\t \delta^{-1/2}}\,dy \,\frac{y}{(1+y^2)^2}\,e^{-i \t k \t \delta^{1/2}(y+1/y)}\,.
\ee
Plugging this into the modular Hamiltonian obtains
\be
\Delta\langle \mc H \rangle=2i \int_{\t \delta^{1/2}}^{\t \delta^{-1/2}}\,dy \,\frac{y}{(1+y^2)^2}\,\int_0^\infty d\t k\,\frac{1}{1+2i \t k/\t m}\,e^{-i \t k \t \delta^{1/2}(y+1/y)}\,.
\ee

Now we perform the integral over $\t k$ in terms of the exponential integral function, finding
\be
\Delta\langle \mc H \rangle=-2 \t m \int_{\t \delta^{1/2}}^{\t \delta^{-1/2}}\,dy \,\frac{y}{(1+y^2)^2}\, e^{\frac{1}{2}\t m \t \delta^{1/2}(y+1/y)}\,\text{Ei}\left(-\frac{1}{2}\t m \t \delta^{1/2}(y+1/y)\right)\,.
\ee
The remaining integral over $y$ may be performed numerically, and exhibits a logarithmic divergence in $\t m \t \delta^{1/2}$. We conclude that
\be
\Delta\langle \mc H \rangle \propto \t m \,\log(\t m \t \delta^{1/2})\,,
\ee
reproducing the result used in the main text.

\bibliographystyle{JHEP}
\renewcommand{\refname}{Bibliography}
\addcontentsline{toc}{section}{Bibliography}
\providecommand{\href}[2]{#2}\begingroup\raggedright

\end{document}